\documentclass[prd,singlecolumn,nofootinbib,showkeys,citeautoscript]{revtex4}
\pdfoutput=1
\usepackage{hyperref}
\usepackage{graphicx}
\usepackage{epstopdf}
\usepackage{amsmath}
\usepackage{dcolumn}
\usepackage{multirow}
\usepackage{fancybox}
\usepackage{bm}
\usepackage{multirow}
\usepackage{array}
\usepackage{rotating}
\usepackage{ulem,fancyvrb}
\usepackage{latexsym}
\usepackage{amsfonts}
\usepackage{xcolor}
\usepackage{longtable}
\usepackage{amssymb}
\def\tana2{\tan^2\alpha_H}
\def\tanb2{\tan^2\beta}

\def\m0pr{m_0'}

\def\Mz2{M_Z^2}

\def\met100{\slashed{E}_T\geq 100~{\rm GeV}}

\newcommand{\beqn}{\begin{eqnarray}}
\newcommand{\eeqn}{\end{eqnarray}}
\newcommand{\be}{\begin{equation}}
\newcommand{\ee}{\end{equation}}

\def \n34{\tilde{\chi}^{0}_{3,4}}

\def\125{$\sim 125$ GeV~}

\def\st{Stueckelberg~mechanism~}

\allowdisplaybreaks[1]

\begin{document}

\title{Higgs Physics and Supersymmetry}

\author{Pran~Nath} 
\email{nath@neu.edu}
\affiliation{Department of Physics, Northeastern University,
 Boston, MA 02115-5005, USA}

\begin{abstract}
A brief overview of  Higgs physics  and of supersymmetry is given.
The central theme of the overview is to explore the implications of the 
recent discovery of a Higgs like particle regarding the prospects for the 
discovery of supersymmetry assuming that it is indeed
the  spin 0 CP even boson that enters in the spontaneous breaking of the
electroweak symmetry. The high mass of the Higgs like boson at $\sim 125$ GeV
points to the weak scale of supersymmetry  that enters in the 
loop correction to the Higgs boson mass, to be relatively high, i.e., in the TeV
region. However, since more than one independent mass scales enter in 
softly broken supersymmetry, the allowed parameter space of supersymmetric
models can allow a small Higgs mixing parameter $\mu$ and light gaugino
masses consistent with a $\sim 125$ GeV Higgs boson mass.  Additionally some light
third generation sfermions, i.e.,  the stop and the stau are also permissible.  
Profile likelihood analysis of a class of SUGRA models indicates that $m_A> 300$ GeV
which implies one is in the decoupling phase and the Higgs couplings are close
to the standard model  in this limit. Thus a sensitive measurement of 
the Higgs couplings with fermions and with the vector bosons is  needed 
 to detect beyond the standard model effects.
  Other topics discussed
include  dark matter, proton stability, and the Stueckelberg 
extended models as probes of new physics. A brief discussion of the way forward 
in the post Higgs discovery era is given. 
\end{abstract}

\maketitle

\section{Introduction}

The Higgs boson~\cite{Englert:1964et,Higgs:1964pj,Guralnik:1964eu} is a central part of the
standard model of electroweak interactions~\cite{Weinberg:1967tq,salam}. 
The long search for it culminated 
with the  observation of a boson in the mass range 125-126 GeV (see Eq.(\ref{atlascms}))
by  the ATLAS ~\cite{:2012gk} and 
 CMS~\cite{:2012gu} Collaborations  using the combined 7~TeV and 8~TeV data 
 (For earlier results from CMS and ATLAS Collaborations which gave a strong indication of a Higgs signal see
 Refs. ~\cite{Khachatryan:2011tk} and ~\cite{Aad:2011hh}).  
 Pending  a more precise determination of its mass we will refer to new particle as $\sim 125$ GeV boson.  
 The properties of it still need to be fully established even though there is a strong belief 
 that the $\sim 125$ GeV boson is indeed the long sought after Higgs boson of electroweak theory.
  Here we will first review the current status of the recent developments. Our focus, however, will
  be on the implications of this discovery for supersymmetry assuming that the newly discovered
  particle is indeed the Higgs boson. Specifically we will show that the $\sim 125$ GeV Higgs boson
  mass is consistent within  SUGRA grand unification~\cite{Chamseddine:1982jx}
  which predict  an upper limit on the Higgs
  boson mass of $\sim 130$ GeV~\cite{Akula:2011aa,Arbey:2012dq,Ellis:2012aa}. 
  A \125 Higgs boson  mass puts severe constraints on the parameter space of SUGRA models and 
  it is shown that most of the parameter space consistent  with
  the  $\sim125$ GeV Higgs mass lies on the Hyperbolic Branch of radiative breaking of the 
  electroweak symmetry, and specifically on Focal Surfaces. The implications of the  $\sim 125$ GeV 
  Higgs mass for dark matter is also discussed. Here it is shown that within SUGRA models with 
  universal boundary conditions most of the parameter space of the model lies between 
  the current sensitivity of XENON-100 and the sensitivity of  future XENON-1T or SuperCDMS-1T.
  Analyses within mSUGRA show ~\cite{Akula:2012kk}  
  that the $\sim 125$ GeV  Higgs mass constraint implies that the CP odd Higgs
  mass $m_A >300$ GeV  (see also Ref.~\cite{Maiani:2012qf}), and one is in the decoupling limit.
      In this limit the deviations of the
  Higgs couplings to matter and to vector fields from the standard model values   
  are expected to be typically small and sensitive 
  measurement of these couplings are needed to discern beyond the standard model effects.
  A brief discussion is given of the two photon decay of the Higgs which appears to show 1-2 $\sigma$ deviation
  from the standard model result. An explanation of such deviations would require some light
  particles  in the loops and there are a variety of models that attempt to accomplish that. 
  We also discuss briefly the idea of natural SUSY promoted recently.

  There are  several other topics discussed in this review. One topic  concerns
  the so called cosmic coincidence which refers to the fact that dark matter to 
  visible matter is in the ratio $\sim 5:1$. We will discuss how the cosmic coincidence can work
  in supersymmetric theories.  We will also discuss SUSY grand unification specifically regarding 
  the solution to the doublet-triplet splitting in $SO(10)$ using the missing partner mechanism.
  Other topics discussed within SUSY grand unification include  the  status of proton lifetime
  in supersymmetric theories and its current limits from experiment
  and the radiative breaking of R parity and how such a breaking can be evaded. 
  We also discuss the (Higgsless)  Stueckelberg extensions of the standard model (SM) gauge group 
  and their implications at colliders. Finally we discuss the prospects for the observation of 
  supersymmetry at the LHC.

  The outline of the rest of the paper is as follows: In Sec. 2 we discuss the recent developments
  regarding the discovery of the $\sim 125$ GeV boson. In Sec.    3  we discuss the implications 
  of this discovery for supersymmetry and  the naturalness of TeV scalars
  on the Hyperbolic Branch of radiative breaking of the electroweak symmetry. The other topics
  discussed in this section are natural SUSY and  
  the decoupling limit. In Sec. 4 we discuss 
  supersymmetry and the cosmic frontier with focus on dark matter. Here we also discuss  
   cosmic coincidence in the context of supersymmetry and show that  it results 
  in a two component dark matter. In Sec. 5 we discuss the recent progress on the doublet-triplet
  splitting in $SO(10)$ SUSY grand unified models.  A review of the 
  current status of proton stability in SUSY grand unification and the current experimental
  limits on proton decay lifetimes is given. Radiative breaking of R parity and the means to avoid 
  such a breaking are also discussed. Sec. 6  describes the Stueckelberg extensions and
  its relation to strings, and possible tests of such models at colliders. Prospects for SUSY 
  in future data from the LHC are discussed in Sec. 7. In Sec. 8 the way forward after the discovery of
  the Higgs is discussed.

\section{Recent Developments} 
The analysis by ATLAS ~\cite{:2012gk} and CMS~\cite{:2012gu} using the combined 7 ~TeV and 8 
~TeV data indicates observation of a  signal for a boson with mass 

\beqn
 {\rm ~~CMS:}   ~~125.3\pm 0.4 ({\rm stat})\pm 0.5({\rm sys})~{\rm GeV}  ~~~~~5.0\sigma, \nonumber\\
{\rm ATLAS:} ~~ 126.0 \pm0.4 ({\rm stat})\pm 0.4({\rm sys})~{\rm GeV}  ~~~~~5.9\sigma. 
\label{atlascms}
\eeqn
The search channels included $h\to \gamma\gamma$, $h\to ZZ^*\to 4{\it l}$, $h\to WW^*\to 2{\it l}2 \nu$
as well $h\to b \bar b, \tau \bar\tau$ with $h\to \gamma\gamma$ being the dominant discovery channel.
One can ask what kind of an object the observed  boson is. The two photon final state indicates that it is not a 
spin one particle due to the Landau-Yang theorem~\cite{Yang:1950rg}.
But it could be spin 0 or spin 2. 
Assuming  it is spin 0 one can ask is it  a CP even or a CP odd particle [The spin and CP properties can be determined by analysis of the processes
$pp\to h^0\to ZZ\to l_1 \bar l_1 l_2 \bar l_2$~\cite{Choi:2002jk}
and $pp\to h^0\to W^+W^- \to l^+l^-\nu\bar \nu$.
 See \cite{Ellis:2012wg} and the references therein.].  
Suppose now that it  is
a CP even particle. In this case, it could be  either an ordinary spin 0 particle or a 
a spin 0 particle that is a remnant of spontaneous breaking  which gives mass to gauge bosons, 
quarks,  and leptons~\cite{Englert:1964et,Higgs:1964pj,Guralnik:1964eu,Guralnik:2009jd}.
Let us assume that the new boson is a remnant of spontaneous breaking. 
One may ask 
what the underlying model is in this case,
i.e., is the underlying model the  standard model, the supersymmetric standard model, a string model,
  a composite Higgs model, or something else altogether. 
Finally assuming that the newly discovered particle  is a SUSY Higgs boson
  one  can ask if its properties can shed some light on the  breaking of supersymmetry
 such as if supersymmetry is broken via  gravity mediation, gauge mediation, anomaly mediation or 
 something else. \\

 Definitive answers to  many of these questions will have to wait for more data. However,    
partial answers already exist in the data from the Tevatron~\cite{:2012zzl}
 and from the CMS/ATLAS experiments\cite{:2012gu,:2012gk}.
For example, there is  evidence already that the  boson that  is discovered 
is a remnant of spontaneous breaking  responsible for giving mass to quarks and leptons.  
  Thus an important test  that the scalar boson is responsible  for generating mass for fermions is 
 $ h_t ({\rm Yukawa}) \sim \frac{m_f}{v}$   where $v$ is the VEV of the scalar field. 
   This means that a heavier fermion has a larger branching ratio than a lighter one in the decay of the Higgs.
    This is consistent with the experimental  data from the Tevatron~\cite{:2012zzl}   and from 
   the    CMS~\cite{:2012gu} and ATLAS ~\cite{:2012gk}   Collaborations 
     where only $b\bar b, \tau\bar \tau $ are seen but e.g., 
   $\mu^+\mu^-$ is not seen. 
  However,  a test of all the couplings will be needed to establish the discovered boson to  be a  
  truly standard model like Higgs  boson,
i.e., a test of the couplings to fermions  and
to vector bosons  specifically the couplings $ hb\bar b, ~ht\bar t, ~h\tau\bar \tau, ~hWW, ~hZZ, ~h \gamma\gamma, ~h  Z\gamma$
(see, e.g.,  Ref. \cite{Gunion:1989we}).\\
 
\begin{figure}[t!]
\begin{center}
\hspace{-.2cm}
\includegraphics[scale=0.15]{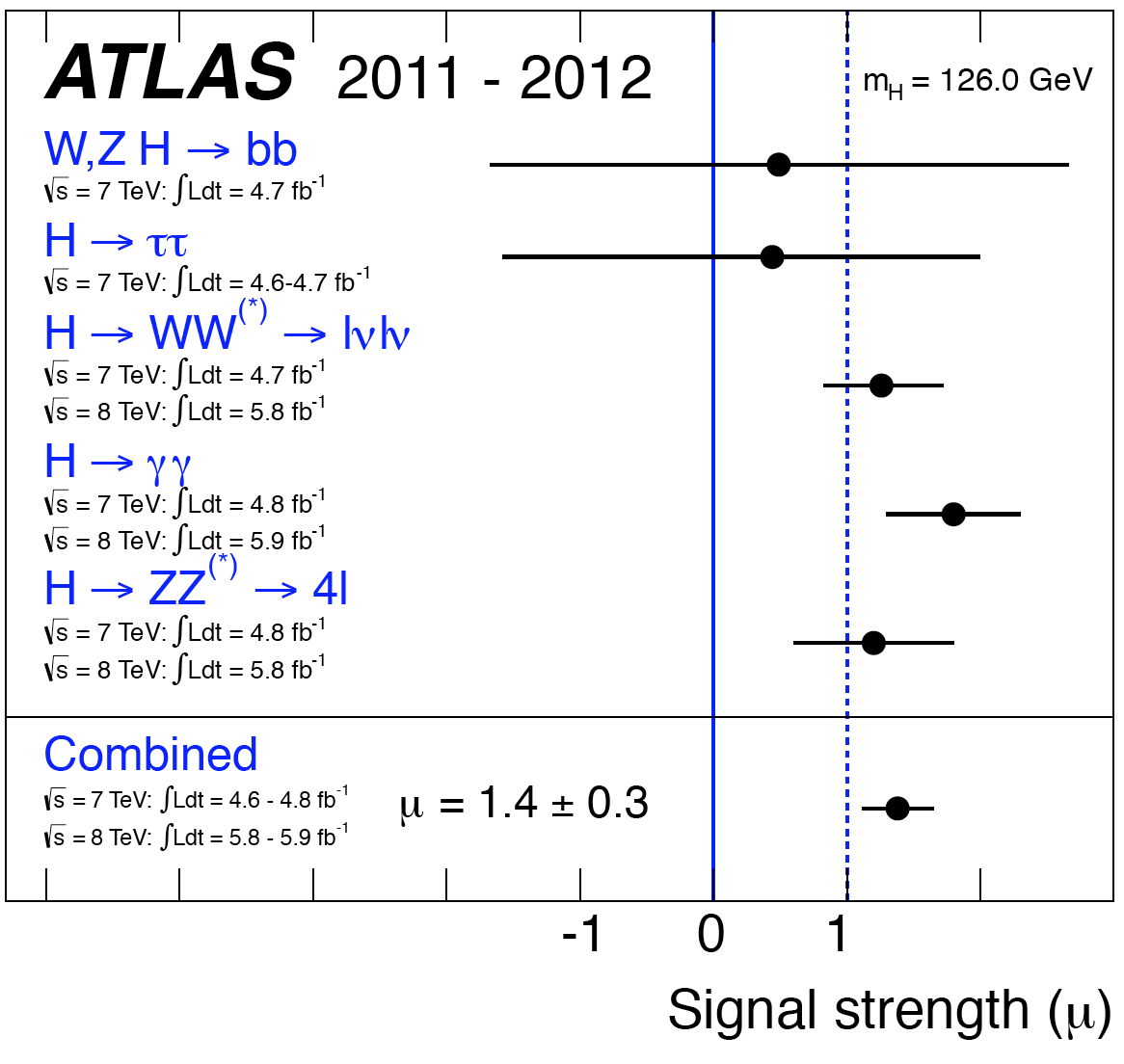}
\hspace{0.4cm}
\includegraphics[scale=0.15]{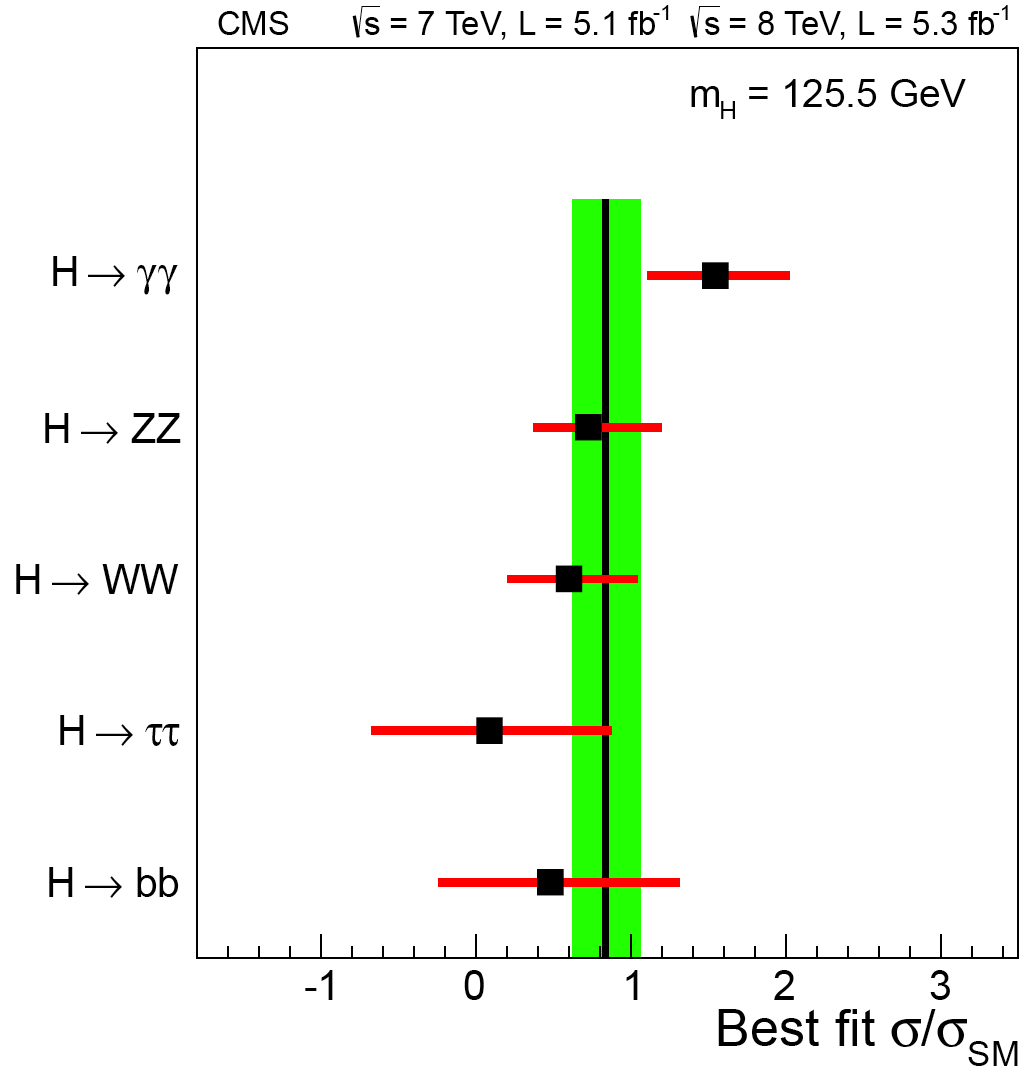}
\caption{The left panel gives the analysis by the ATLAS Collaboration ~\cite{:2012gk} for the signal strength
parameter $\mu$. 
The right panel gives the  analysis by the CMS Collaboration~\cite{:2012gu} where 
 $\sigma/\sigma_{SM}$ is plotted,
  where $\sigma$ refers to the production cross section times the relevant branching fractions while $\sigma_{SM}$ is the expectation for the SM case.    Statistical and 
systematic uncertainties on the evaluation of $\sigma/\sigma_{SM}$  are indicated by  horizontal
bars which correspond to $\pm 1$ deviation uncertainties. }
 \label{fig1}
\end{center}
\end{figure}

 Important clues to new physics can emerge by looking at deviations of  the Higgs boson couplings from the standard model predictions.  
 In the left panel of Fig.(\ref{fig1})  the signal strength ($\mu$) as given by the 
 ATLAS Collaboration~\cite{:2012gk} is presented  where  $\mu$ is defined~\cite{:2012gk} so that $\mu=0$ corresponds to just the background and $\mu=1$ when one includes the contribution from the standard model Higgs in addition to the background for $m_h=126$ GeV  for various decay final states for the observed boson. 
The right panel of Fig.(\ref{fig1}) gives the  analysis by the CMS Collaboration~\cite{:2012gu}.
Here $\sigma/\sigma_{SM}$ is plotted where $\sigma$ refers to the production cross section times the relevant branching fractions while $\sigma_{SM}$ is the expectation for the SM case. 
The possibility that the CMS and ATLAS  data  may have  an excess in the diphoton channel has 
already led to   some excitement. Thus the CMS and ATLAS Collaborations give~\cite{:2012gu,:2012gk}

\beqn
  \hat \mu \equiv \frac{\sigma(pp\to h)_{obs}}{\sigma(pp\to h)_{SM}} 
= 1.4\pm 0.3 {\rm ~(ATLAS)}, ~0.87 \pm 0.23 {\rm ~(CMS)} \, ,
\eeqn

\beqn
 R_{\gamma \gamma} \equiv \hat \mu \frac{\Gamma(h\to\gamma\gamma)_{obs}}{ 
 \Gamma(h\to \gamma\gamma))_{SM}} 
= 1.8\pm  0.5 ~({\rm ATLAS }), ~1.6\pm 0.4~({\rm CMS}) \, .
\eeqn
In the SM the Higgs boson can decay into photons via $W^+W^-$ and $t\bar t$  in the loops.
The W-loop gives the larger contribution which is partially cancelled by the top-loop.
To enhance the decay one could add extra particles in the loops (vectors, fermions, scalars) and various works  have appeared
along these lines in the literature. Specifically, one possibility considered is light particles, e.g., 
staus or light  stops,
in the loops
~\cite{Carena:2011aa,Carena:2012xa,Giudice:2012pf}.
One may parametrize the diphoton rate as follows 
\beqn
R_{\gamma \gamma}
  =  \hat \mu \left[ -1.28 (1+ \delta_V)
  + 0.28 (1+ \delta_f)  + \delta_S\right]^2. 
  \eeqn
  The $\delta's$ stand for deviations from the standard model prediction
  (For a sample of models where such deviations from the standard model are discussed 
  see Refs.~\cite{Joglekar:2012vc,ArkaniHamed:2012kq,An:2012vp,Abe:2012fb,Chang:2012tb}). 
 An alternative solution is that the excess seen in the diphoton channel is simply
a result of QCD uncertainties~\cite{Baglio:2012fu}.

\section{Implications for supersymmetry}
We discuss now the implications of the recent  discovery~\cite{:2012gu,:2012gk}  for supersymmetry . 
From here on we will assume that the discovered boson is indeed the Higgs boson.
Now prior to the LHC results of Refs. ~\cite{:2012gu,:2012gk}  it was generally thought 
that the Higgs  boson mass could  lie between the lower limit given by LEP of 115 GeV and the upper limit 
which is constrained  by partial wave unitarity to be  around 800 GeV~\cite{Dicus:1992vj,Lee:1977yc}.
Inclusion of vacuum stability gives additional more stringent constraints.

\begin{figure}[t!]
\begin{center}
\hspace{-.2cm}
\includegraphics[scale=0.3]{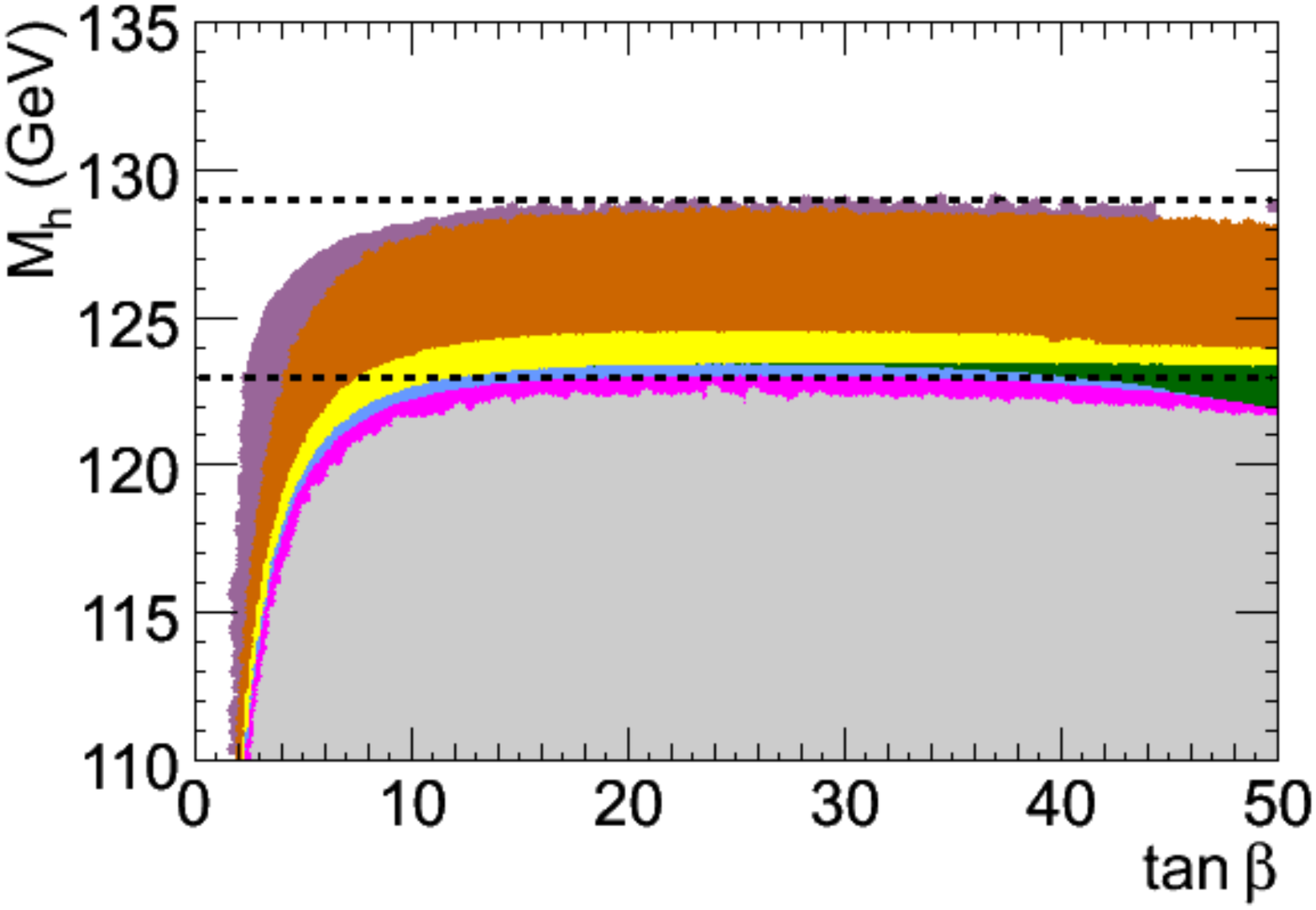}
\hspace{-1.5cm}
\includegraphics[scale=0.225]{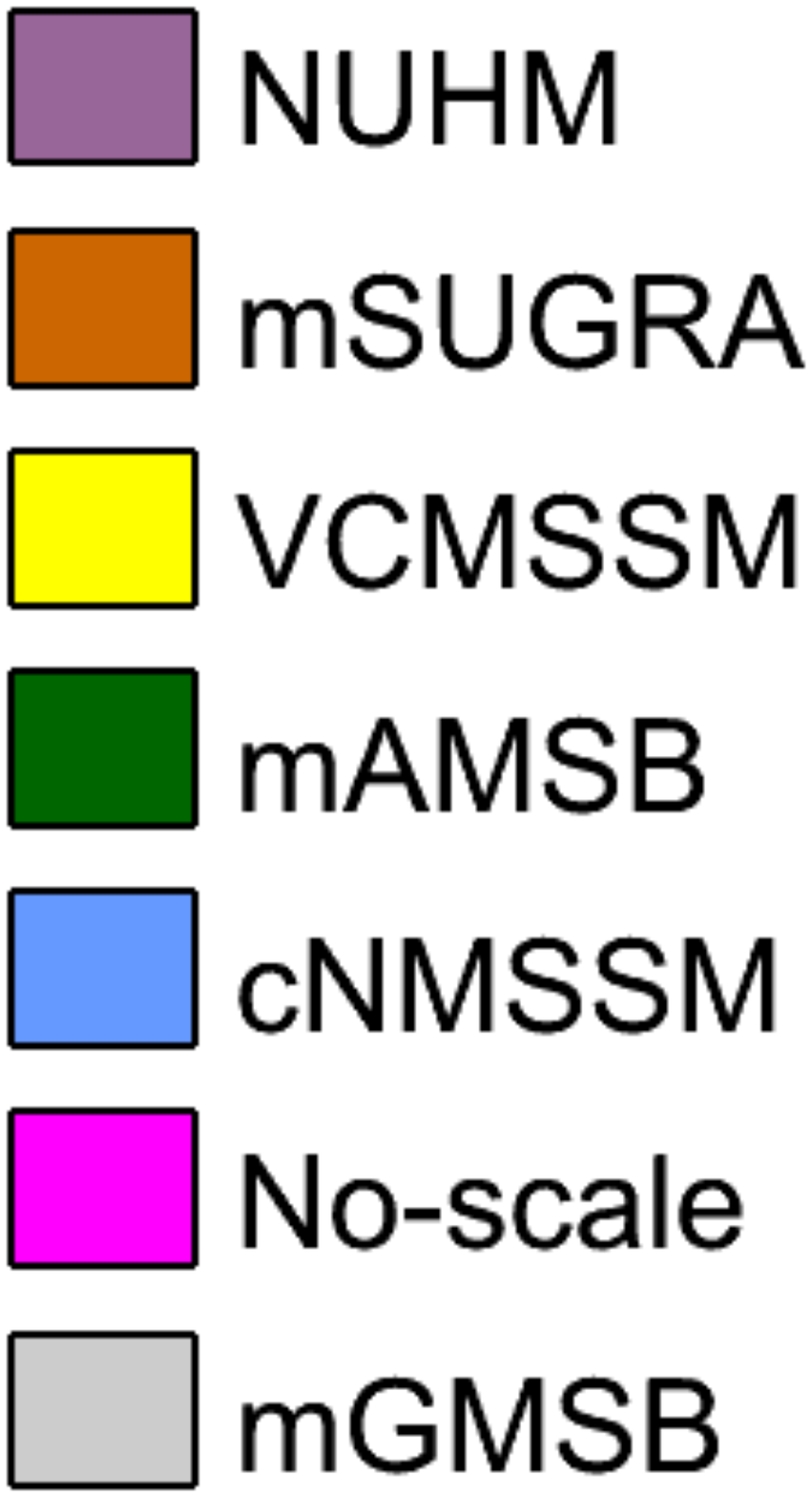} 
\caption{An analysis of the ranges of the Higgs boson mass vs $\tan\beta$ 
for a variety of different  models as given in Arbey et.al.~\cite{Arbey:2012dq,Arbey:2011ab}.
The upper limit on $M_S$ is chosen to  be 3 TeV. 
The figure  is taken from Ref.~\cite{Arbey:2012dq}.}
 \label{fig2}
\end{center}
\end{figure}

The analysis of the vacuum stability is  sensitive 
to corrections beyond the leading order and a recent  analysis at  
the next-to-next -leading order (NNLO) requiring that the vacuum be absolutely stable 
 up to the Planck scale gives ~\cite{Degrassi:2012ry} 
\beqn m_h 
> 129.4 + 1.4 \left( \frac{m_t -173.1}{0.7} \right)
-0.5\left( \frac{\alpha_s(M_Z)-0.1184}{0.0007}\right) \pm 1.0\ .
\label{stability}
\eeqn 
where  $m_h$  and $m_t$ are in GeV.
With the inclusion of both the theoretical error in the evaluation of $m_h$  
estimated at  $\pm 1.0$ GeV and the experimental
errors on the top mass and $\alpha_s$ the analysis of Ref. \cite{Degrassi:2012ry} gives
\beqn m_h > 129.4 \pm 1.8~{\rm GeV},
\eeqn 
 for the standard model to have vacuum stability  up to the Planck scale.
This excludes the vacuum stability for 
the SM for  $m_{h^0} < 126$ GeV at the $2\sigma$ level~\cite{Degrassi:2012ry}. Thus the observation 
of a Higgs mass of $\sim 125$ GeV would give vacuum stability up to only scales between $10^{9}- 10^{10}$ GeV
and stability up to the Planck scale would require new physics (see, however, Refs. \cite{Alekhin:2012py,Masina:2012tz}).
Such new physics could be 
supersymmetry but  other models can be found that can also achieve such a  result (for  recent examples see
    Refs~\cite{Chamseddine:2012sw,Anchordoqui:2012fq}). 
However, there  still remains  the well known hierarchy problem,
i.e., the standard model gives large loop corrections to the Higgs boson mass 
\beqn
 m^2_h = m_0^2 + O(\Lambda^2) \, .
\eeqn
Since  $\Lambda \sim O(10^{16})$ GeV  a large fine tuning is involved to get the Higgs mass in the electroweak region.
The SUSY version of SM is  free of the large fine tuning problems. Thus MSSM 
has no large hierarchy problem and gives perturbative physics up to the GUT scale. 
  In SUSY the Higgs boson mass at tree level is $< M_Z$~\cite{Nath:1983fp}
    and loop corrections are needed
  to lift it above
$M_Z$.The dominant one loop contribution arises from the top-stop sector and is given by
\cite{Carena:2002es,Djouadi:2005gj}

\beqn
\Delta m_{h^0}^2\simeq  \frac{3m_t^4}{2\pi^2 v^2} \ln \frac{M_{\rm S}^2}{m_t^2} 
+ \frac{3 m_t^4}{2 \pi^2 v^2}  \left(\frac{X_t^2}{M_{\rm S}^2} - \frac{X_t^4}{12 M_{\rm S}^4}\right)+\cdots~,
\label{deltamh}
\eeqn
where  $v=246$~GeV ($v$ is the Higgs VEV), $M_{\rm S}$ is an average stop mass, and $X_t$ is given  by
  \beqn
 X_t\equiv A_t - \mu \cot\beta~.  
 \eeqn
  The loop correction is maximized  when   
   $ X_t \sim \sqrt 6 M_{\rm S}$.  
It should be noted that when we  talk  about  supersymmetry we are actually taking about local 
 supersymmetry~\cite{Nath:1975nj,Arnowitt:1975xg}
   or supergravity~\cite {Freedman:1976xh,Deser:1976eh,Nath:1976ci}
   since model building makes sense only in such a framework.  
 This is so because breaking of supersymmetry would require
 the cancellation of the vacuum energy.  This can happen in models based on supergravity
because the scalar potential is not positive definite and it is possible to fine tune the vacuum 
energy to zero. Alternately there may be stringy mechanisms for the vacuum energy to be 
 be small~\cite{Sumitomo:2012wa}.
 Further, it should be noted that since supergravity is the field point limit of strings
 an analysis  within the framework of supergravity
 can provide a wide class  of models including those arising from strings (such as, e.g.,~\cite{Acharya:2008zi,Conlon:2005ki}) .
 Supergravity grand unified models~\cite{Chamseddine:1982jx,Nath:1983aw} provide a reliable
 framework for the exploration of physics from the electroweak scale up to the GUT scale of $\sim 10^{16}$
 GeV.  Under the assumption of  universal boundary conditions the soft breaking sector of this 
 model gives a very constrained  set of soft terms ~\cite{Chamseddine:1982jx,Nath:1983aw,Hall:1983iz}
 (for a review see Ref. \cite{Nath:2003zs}). 
Specifically  the soft breaking sector of this theory  after radiative breaking of the electroweak symmetry is given by the set of
    parameters  $m_0, m_{1/2}, A_0, \tan\beta$ and the sign of the Higgs mixing parameter $\mu$.
    In this model usually referred to as mSUGRA or CMSSM,  the upper limit on the Higgs mass     
    is $m_h\sim 130$ ~GeV~\cite{Akula:2011aa,Arbey:2012dq,Ellis:2012aa}. 
    This result is to be contrasted with the more general result in SUSY models where one expects the Higgs mass to lie below $\sim 150$ GeV ~\cite{Kane:1992kq}. 
The upper limit predicted in mSUGRA is exhibited in  Fig. (\ref{fig2}). 
It is meaningful that at the end of the day the Higgs mass ended up below the upper limit predicted in mSUGRA. 
This provides a positive signal that what we are seeing is a  SUSY Higgs. 
The implications of the new results ~\cite{:2012gu,:2012gk} 
have been analyzed in a number of works
\cite{Cao:2012yn,Cohen:2012wg,Abe:2012fb,Cheng:2012pe,Dev:2012ru,Ajaib:2012eb,
Kaplan:2012zzb,Liu:2012qu,Ibanez:2012zg,Hebecker:2012qp,
Aparicio:2012iw,Li:2012jf,Boehm:2012rh,Ellis:2012xd,CahillRowley:2012rv,Feng:2012jf,Perez:2012gg,Dhuria:2012bc,Wymant:2012zp} .

\begin{figure}[t!]
\begin{center}
\hspace{-1.0cm}
\includegraphics[scale=0.18]{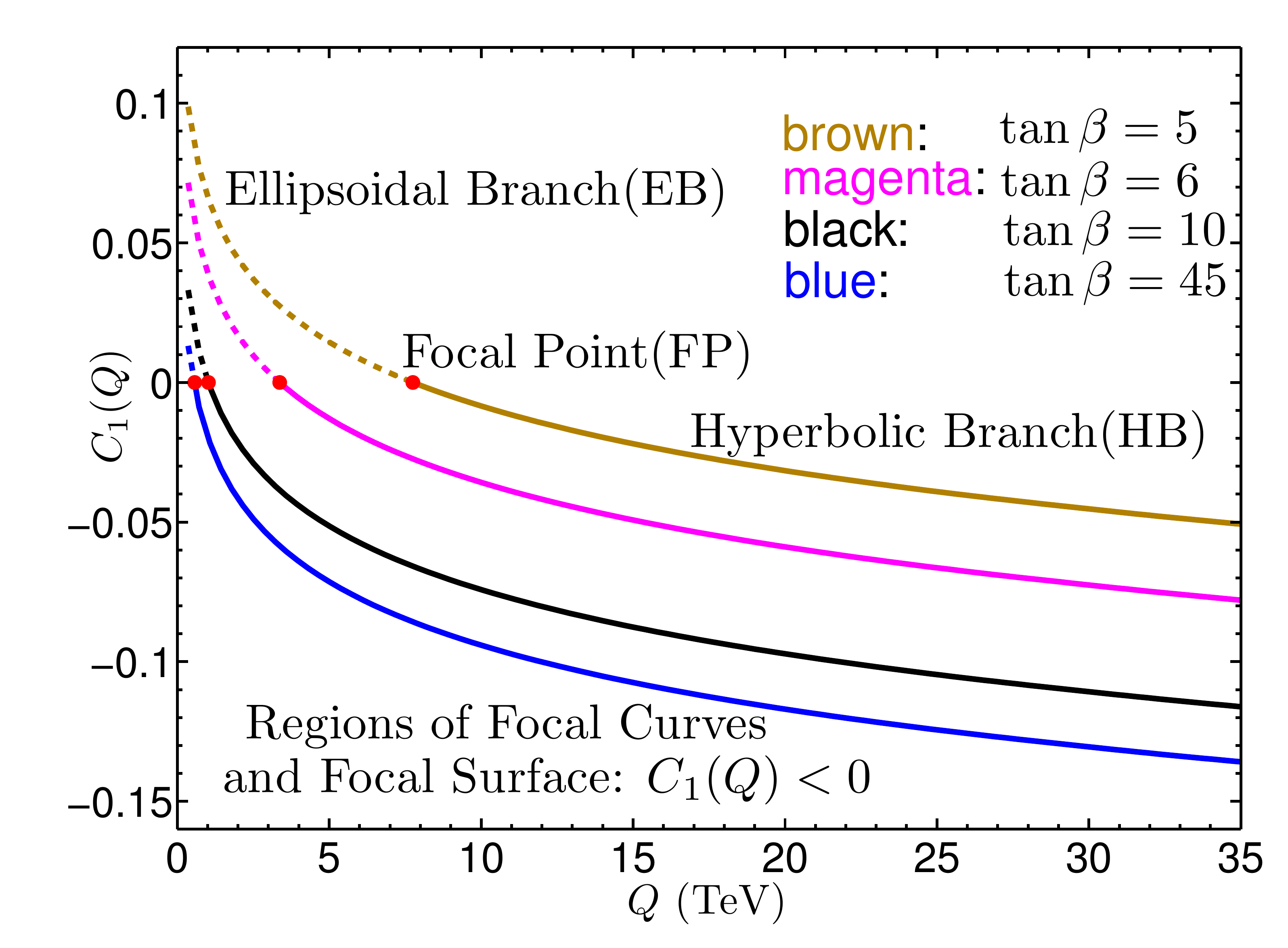}
\includegraphics[scale=0.24]{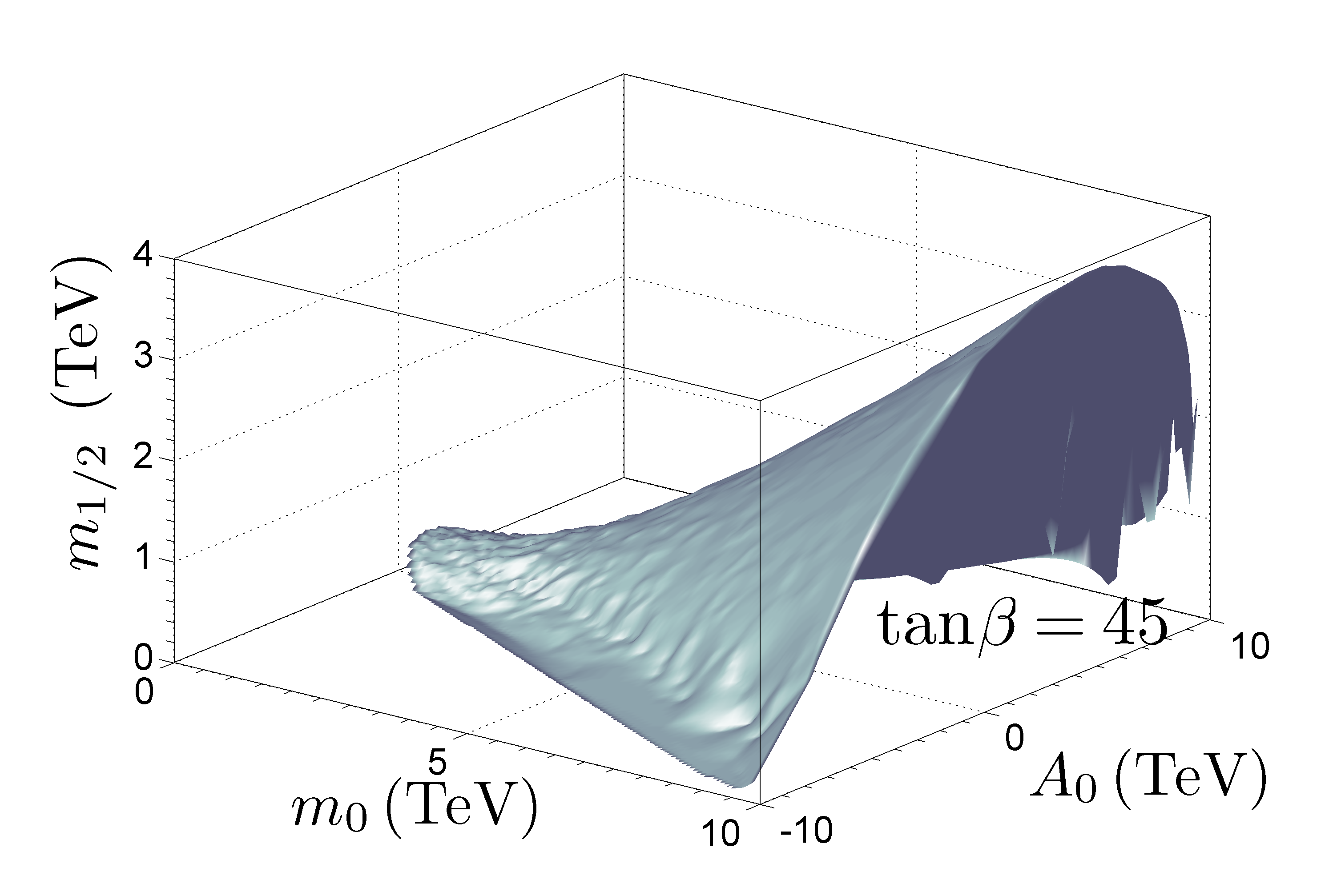}
\caption{Left panel:  $C_1(Q)$  as a function of $Q$ 
 for   $\tan\beta=5$~(brown), $\tan\beta=6$~(magenta), $\tan\beta=10$~(black) and $\tan\beta=45$~(blue). 
Right panel:  A Focal Surface for the case and  $\tan\beta =45$   while 
$m_0, m_{1/2}, A_0$ can  all get large while $\mu$ remains fixed so that 
$\mu=\left(0.465\pm0.035\right)$ TeV. Figs. taken from Ref. ~\cite{Akula:2011jx}.}
 \label{fig3}
\end{center}
\end{figure}

\subsection{Natural  TeV Size Scalars}
 TeV size scalars arise naturally in SUGRA models with radiative breaking of the electroweak symmetry.
Thus the radiative breaking of the electroweak symmetry allows a determination of the Higgs mixing 
parameter $|\mu|$ such that  
\beqn 
\mu^2 = -\frac{1}{2} M_Z^2 + \frac{ (m_{H_1}^2+ \Sigma_1)- (m_{H_2}^2 + \Sigma_2)\tan^2\beta}
{(\tan^2\beta -1)}  \, , \nonumber\\
\Sigma_a= v_a^{-1} \frac{\partial \Delta V}{\partial v_a} \, , ~a=1,2 \, ,
\label{rewsb}
\eeqn
where $v_a=<H_a>$ and $H_2(H_1)$ give masses to the up quarks (down quarks and leptons) and $\Sigma_a (a=1,2)$ 
are radiative corrections at the one loop level~\cite{Arnowitt:1992qp}. To display the dependence 
of $\mu^2$ on the soft parameters more explicitly one may  write  Eq.(\ref{rewsb}) so that
\beqn
 \mu^2  + \frac{1}{2}M_Z^2  = m^2_0  C_1+  A^2_0 C_2 +
 m^2_{\frac{1}{2}} C_3+  m_{\frac{1}{2}}A_0 C_4+ \Delta \mu^2_{loop} \, , 
 \label{rewsb1}
 \eeqn
where $C_1-C_4$  depend on $\tan\beta$ but do not have an explicit dependence on the
soft parameters $m_0, m_{1/2}, A_0$, while $\mu^2_{loop}$ has a complicated  dependence on
soft parameters since it arises from a loop contribution.   One may write Eq.(\ref{rewsb1}) as 
\beqn
 \mu^2  + \frac{1}{2}M_Z^2  = m^2_0  C_1+ \bar A^2_0 C_2 +
 m^2_{\frac{1}{2}} \bar C_3+ \Delta \mu^2_{loop} \, , 
\label{rewsb2}
 \eeqn
 where $\bar A_0$ and $\bar C_3$ are simply related to $A_0, m_{1/2}, C_2, C_3,C_4$,
 so that   $\bar A_0=A_0 + (C_4/2C_2) m_{1/2}$ and $\bar C_3= C_3 - C_4^2/(4C_2)$.
 The functions $C_1-C_4$ and $\Delta_{loop}^2$ are functions of the renormalization group scale $Q$,
 and the constraint of Eq.(\ref{rewsb2}) become more transparent if one goes to the scale $Q_0$ where
 the loop correction $\Delta_{loop}^2$ is minimized so one may examine the remaining terms to 
 elucidate the constraint. Here one typically finds that $C_2, \bar C_3$ are always positive, but 
 $C_1$ could  be either positive, vanishing or negative. These cases lead to drastically different 
 constraints on the allowed parameter space  due to radiative breaking of the electroweak symmetry.\\
  
  The case  $C_1>0$ implies that the soft parameters lie on the surface of an ellipsoid for fixed 
  $\mu$ and so this branch of radiative breaking  of the electroweak symmetry   
  may appropriately be called the 
  Ellipsoidal Branch (EB)~\cite{Chan:1997bi}. 
 In this case none of the soft parameters can get large for a fixed $\mu$.  
 The case where
 $C_1\leq 0$  will be  called the Hyperbolic Branch~\cite{Chan:1997bi,Chattopadhyay:2003xi,Baer:2012uy}.
 Here one, two or more soft  parameters can get large while $\mu$ can remain small.  
 The case when $C_1=0$ is the one where one soft parameter can get large, i.e., $m_0$ with no
 visible effect on $\mu$ as  can be seen from Eq.(\ref{rewsb2}). 
 One may label this  as the Focal Point  (HB/FP) (it is closely related to the Focus Point of
 Ref. \cite{Feng:1999mn})  and it is  in fact, the transition point  between the branches
 EB and HB.  The case when $C_1<0$ and two parameters can get large while $\mu$ remains fixed
 constitutes the Focal Curve region of the Hyperbolic Branch.  
There are actually two types of Focal Curves: those where $m_0, \bar A_0$ get large
 which we label as HB/FC1 and those where  $m_0, m_{\frac{1}{2}}$ get large with $\mu$ fixed which we label as
 HB/FC2. Focal Curves HB/FC1 typically have the asymptotic  
 property ~\cite{Feldman:2011ud}that $|\bar A_0/m_0| \simeq 1$ when $\mu$ is fixed and $m_{1/2}$ is 
 relatively small compared to $m_0$.
 The case where all the dimensioned soft parameters, i.e., $m_0, m_{1/2}, \bar A_0$,
 get large constitutes the Focal Surface region of the Hyperbolic Branch. Thus the Hyperbolic 
 Branch contains  Focal Points, Focal Curves and Focal Surfaces. 
The left panel of Fig.(\ref{fig3}) gives a pictorial classification of the Ellipsoidal Branch, and of the Hyperbolic 
Branch. Plotted in Fig.(\ref{fig3}) is $C_1(Q)$ as a function of $Q$. The dotted curves identify the 
Ellipsoidal Branch region, and the solid curves the Hyperbolic Branch region while the dots 
are the Focal Points which form the boundary between the Ellipsoidal Branch and the 
Hyperbolic Branch.  
The right panel of Fig.(\ref{fig3}) gives a 
display of a Focal Surface which exhibits regions of the parameter space of mSUGRA where two or 
more soft  parameters among $m_0, m_{1/2}, A_0$ get large while $\mu$ remains  small. 
In the top left panel of Fig. (\ref{fig4}) we exhibit the  allowed parameter space of mSUGRA under the current experimental
constraints. Here one finds  that  most of the parameter space of  mSUGRA lies on focal surfaces which, of course,
include focal curves.

\subsection{Natural SUSY}

Often  a criterion of fine tuning is imposed to discriminate among models. A conventional choice is 
~\cite{Barbieri:1987fn}
\beqn
f =   {\rm max}  \frac{\partial ln m_{h^0}^2}{ \partial ln \alpha_i} 
\label{f}
\eeqn
where $\alpha_i$ are the basic parameters on which $m_{h^0}^2$ depends. 
In the decoupling limit in MSSM on has  
\beqn
m_{h^0}^2= M_Z^2 \cos^22\beta + \Delta m_{h^0}^2
\eeqn
where $\Delta m_{h^0}^2$ is given by Eq.(\ref{deltamh}). For large $tan\beta$ the loop correction 
needed is rather
substantial to raise the Higgs boson mass  from the tree level value  to the experimentally observed value of   $m_{h^0}   \sim 125$ GeV.
In MSSM the scale $M_s$  is in the TeV region implying a large 
average stop mass,  and Eq.(\ref{f}) then leads to a fine 
tuning parameter $f$ of size $O(100)$ or larger.  The level of  fine tuning could  be reduced if there
are additional sources to the  Higgs boson 
mass term which might arise from F term quartics
(see, e.g., Ref.~\cite{Nath:1983fp,Drees:1988fc,Ellwanger:2009dp})  or D term quartics 
(see, e.g., Refs. \cite{Batra:2003nj,Maloney:2004rc}).
In this case a smaller loop correction would be needed leading to a smaller level of fine tuning. 
As an example the F term quartics arise in NMSSM where one includes a singlet field S and a  coupling among the doublets 
and  the singlet field of the form $W\supset  \lambda S H_1H_2$. Here at the 
tree level Higgs mass$^2$ contains an additional term~\cite{Ellwanger:2009dp}
 $\lambda^2 v^2 \sin^22\beta$.
A significant contribution from this sector requires  a large $\lambda$ and a  small $\tan\beta$. However, $\lambda$ is constrained to 
lie below  0.7 otherwise the scale above which non-perturbativity sets is lies below the 
grand unification scale.  In this set up the  stop masses can lie in the few hundred GeV region
and the fine tuning is significantly reduced~\cite{Hall:2011aa}. 
However, the downside of 
including a singlet  is well known (see, e.g., Refs.~\cite{Nilles:1982mp,Bagger:1993ji}).
 A variation of this idea  is $\lambda$SUSY~\cite{Hall:2011aa}. 
Here one considers $\lambda$ larger than 0.7 since that would further
reduce the fine tuning  and values of $\lambda$ as large as 2 are considered. 
The upper limit $\lambda=2$ is taken so that the cutoff scale above which non-perturbativity sets
is does not lie below 10 TeV so  not to disturb  precision electroweak
physics.
 There are several other versions of natural SUSY 
  (see, e.g.,~\cite{Randall:2012dm,Brust:2012uf}).
 Mostly  these are effective theories valid up to around 10 TeV and are not necessarily UV complete.\\

A more generic statement of
natural SUSY is that the particles
 that enter in the solution to the hierarchy problem for the Higgs should be light while the others could be heavy. Thus the particles required to be light are

\beqn 
(\tilde t_L, \tilde b_L), \tilde t_R, \tilde H_2, \tilde H_1 \,
\eeqn
 along with the Higgses.
It is possible that very low mass  stops could have been missed in the LHC analyses and thus it is 
necessary to pursue this possibility to close any existing holes in  signature analyses  for supersymmetry
(for a broad search strategy for natural SUSY see, Refs.~\cite{Essig:2011qg,Allanach:2012vj}).
The current experimental constraints on the stop mass 
from the ATLAS analysis  are exhibited in Fig.(\ref{fig5}) in the neutralino
vs stop mass plane
(see also Ref.
~\cite{:2012si}).
The white regions in Fig.(\ref{fig5}) 
are the ones not excluded by the current data.
Specifically one finds that there is a white region around stop mass of 200 GeV which is yet not excluded by
the experimental constraints. It is hoped that this region will  be probed by the additional data
which will soon be forthcoming  from 
the ATLAS and from the CMS detectors at $\sqrt s=8$ TeV.  Exclusion of the remaining low stop mass region 
 will make  the natural SUSY idea not sustainable.\\

 It should be noted that the choice of criteria one uses to quantify naturalness in the context of electroweak 
 physics are rather subjective and different choices can lead to widely different results. 
 Since the point of contact between theory and the LHC is the existence of light particles which can be
 searched at the LHC, a 
   pragmatic approach is to simply explore the allowed parameter space of models which 
 would lead to light particles consistent with a $\sim 125$ GeV Higgs boson mass constraint without 
 prejudice. 
 In the framework of high scale models, one could accomplish this in a variety of ways.
 For instance, with universal  boundary conditions one can have a   $\sim 125$ GeV 
 Higgs boson mass with large $m_0$ 
 but light gaugino masses with a relatively small $m_{1/2}$ and some light third generation particles 
 can arise with a large trilinear coupling. The other possibility is use of non-universalities both
 in the scalar sector (see, e.g., Refs. \cite{Nath:1997qm,Baer:2012up}) 
 and also in the gaugino sector  (see, e.g., Ref. \cite{Corsetti:2000yq}) to look for a light SUSY mass 
 spectrum. (For a recent work in a top-down approach which examines the range of GUT parameters
 which leads to a natural SUSY spectrum see Ref.~\cite{Baer:2012up}).
  Here we mention that solutions of this type 
 arise on the Hyperbolic Branch (on Focal Points, Focal Curves and 
Focal Surfaces)
as discussed earlier in this section. 
 In these regions $m_0$ is typically large, which has the added benefit that a large $m_0$  helps suppress SUSY FCNC and SUSY CP problems along with the fact that large scalar masses help in stabilizing the 
 proton from lepton and baryon number violating dimension five operators.
 It has  also been proposed that a more realistic way to look at fine tuning is to consider not only the 
  the  EWSB fine tuning but a composite fine tuning parameter which also includes the 
  fine tuning from other sectors as well such as the flavor constraints~\cite{Jaeckel:2012aq}. \\

\subsection{SUSY Higgs in the decoupling phase}
\begin{figure}[t!]
\begin{center}
\includegraphics[scale=0.25]{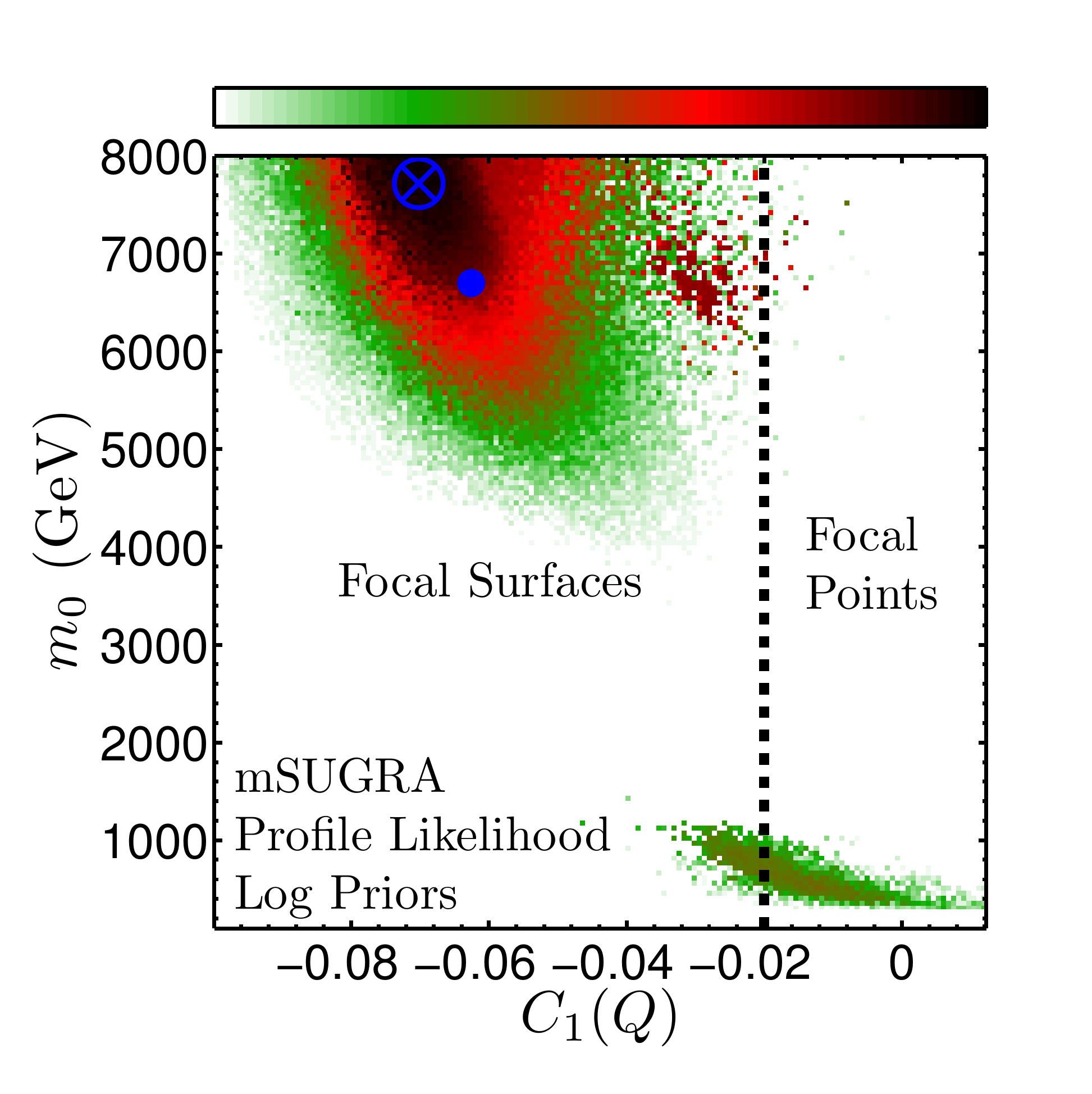}
\hspace{-0.5cm}
\includegraphics[scale=0.27]{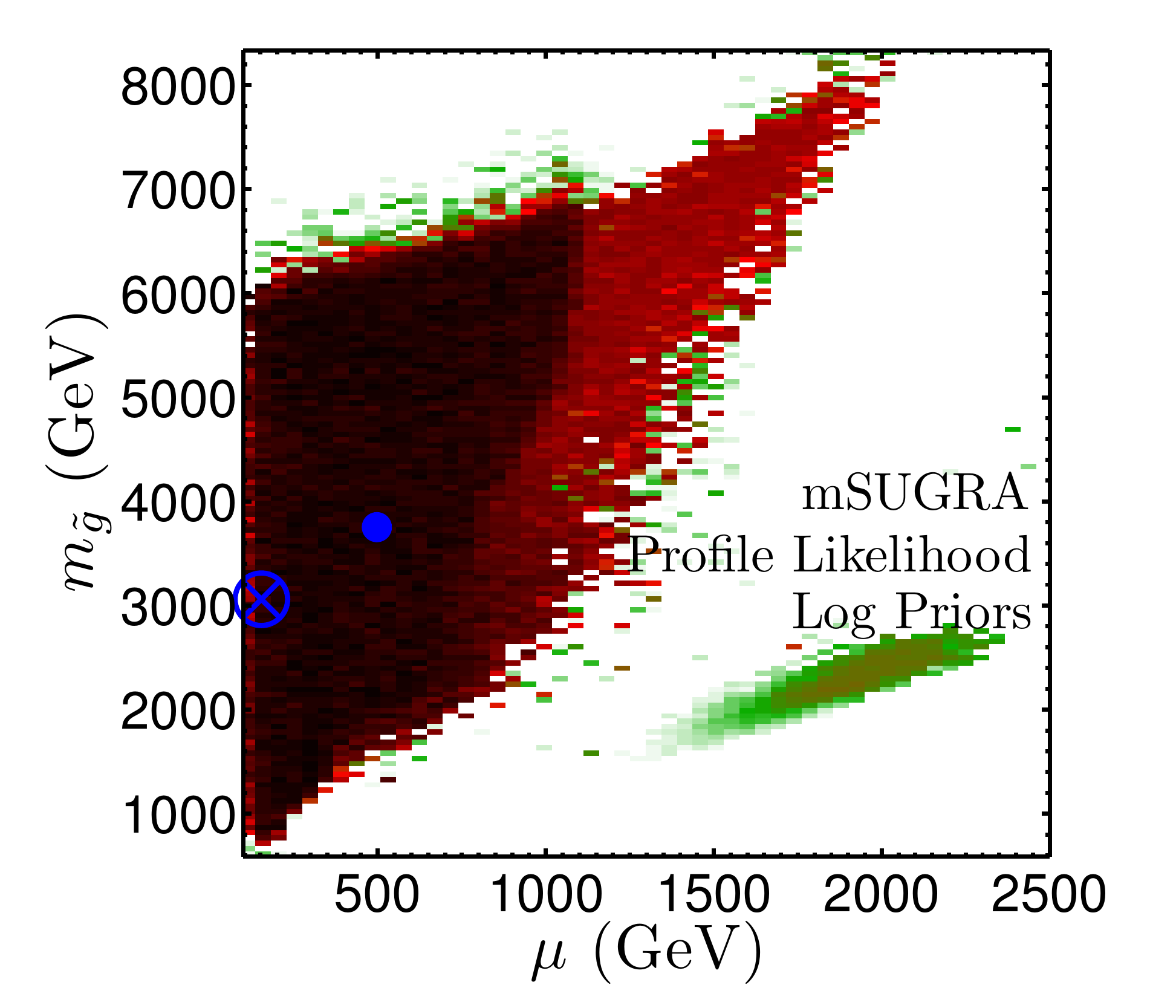}\\
\hspace{-0.2cm}
\includegraphics[scale=0.65]{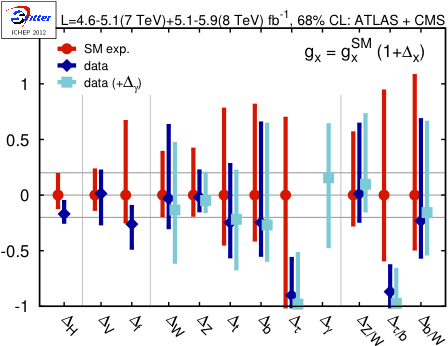}
\hspace{0.4cm}
\includegraphics[scale=0.09]{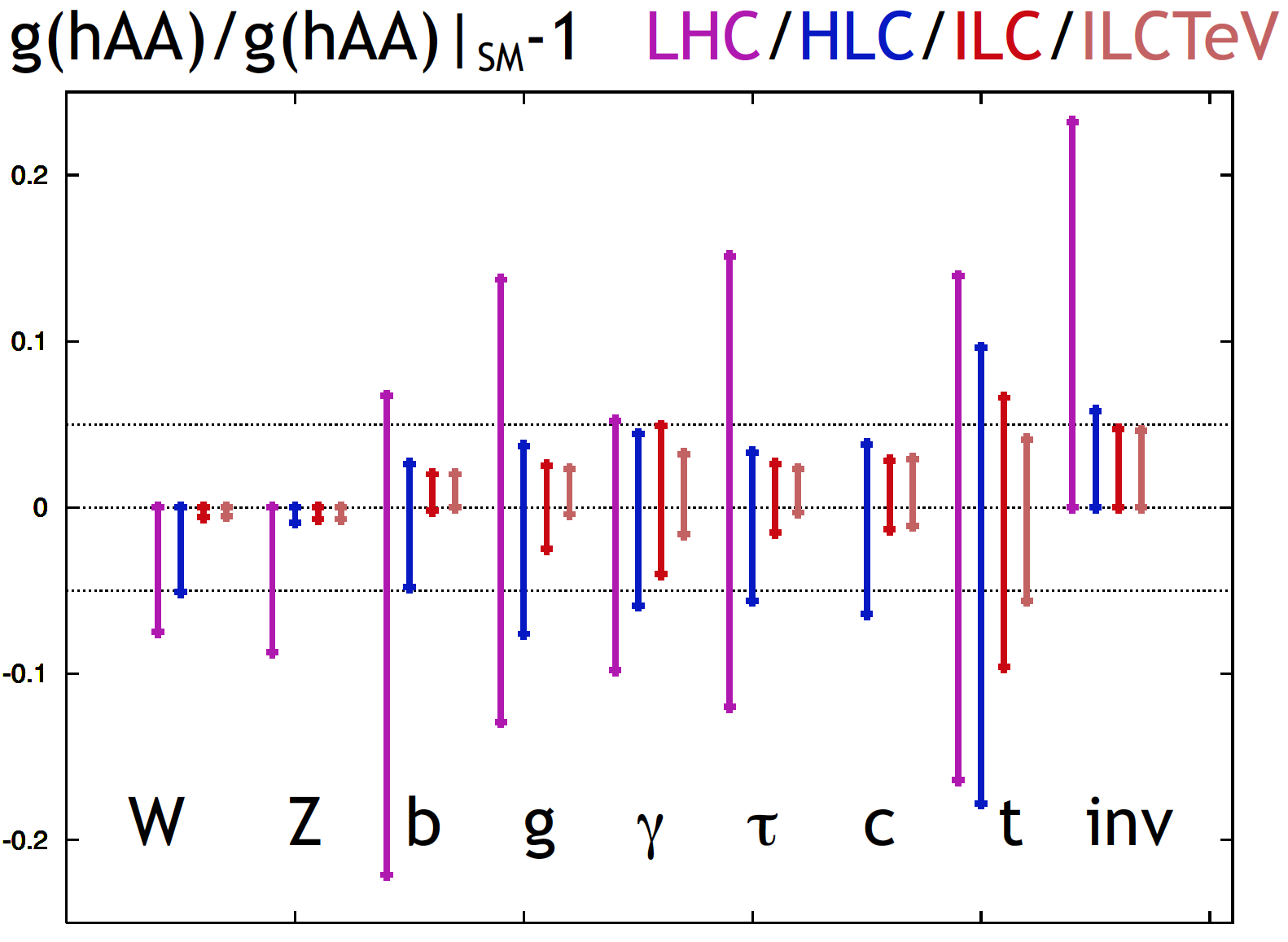}
\caption{
Top left: A profile likelihood analysis of the parameter space of mSUGRA model which exhibits the dispersion in 
$m_0$ as a function of $C_1(Q)$ . The analysis shows that the most of the allowed parameter space consistent with
all constraints lies on focal surfaces. The figure is taken from Akula et.al. ~\cite{Akulax}.
Top right: A profile likelihood analysis in the gluino mass vs the Higgs mixing parameter 
$\mu$.
The analysis shows that there are regions with small $\mu$, which may be construed as natural, where the  
gluino can also be relatively light. The figure is taken from Akula et.al.~\cite{Akula:2012kk}.
Bottom left: Extraction of the Higgs boson couplings using the LHC data. The analysis shows no 
significant deviations from the standard model prediction. However,  the extraction also shows significant uncertainties.
The figure is taken from Plehn et.al. ~\cite{Plehn:2012iz}.
Bottom right: A comparison of the accuracy 
with which LHC  can determine the Higgs couplings  to quarks and leptons and to vector bosons
at  $\sqrt s=14$ TeV and ${\it L} = 300$ fb$^{-1}$ and the accuracy with which ILCTeV can determine
the couplings at $\sqrt s= 1$ TeV and $1000$ fb$^{-1}$ of integrated luminosity.
The figure of this panel is  taken from Peskin~\cite{Peskin:2012we}.}
 \label{fig4}
\end{center}
\end{figure}

\begin{figure}[t!]
\begin{center}
\includegraphics[scale=1.1]{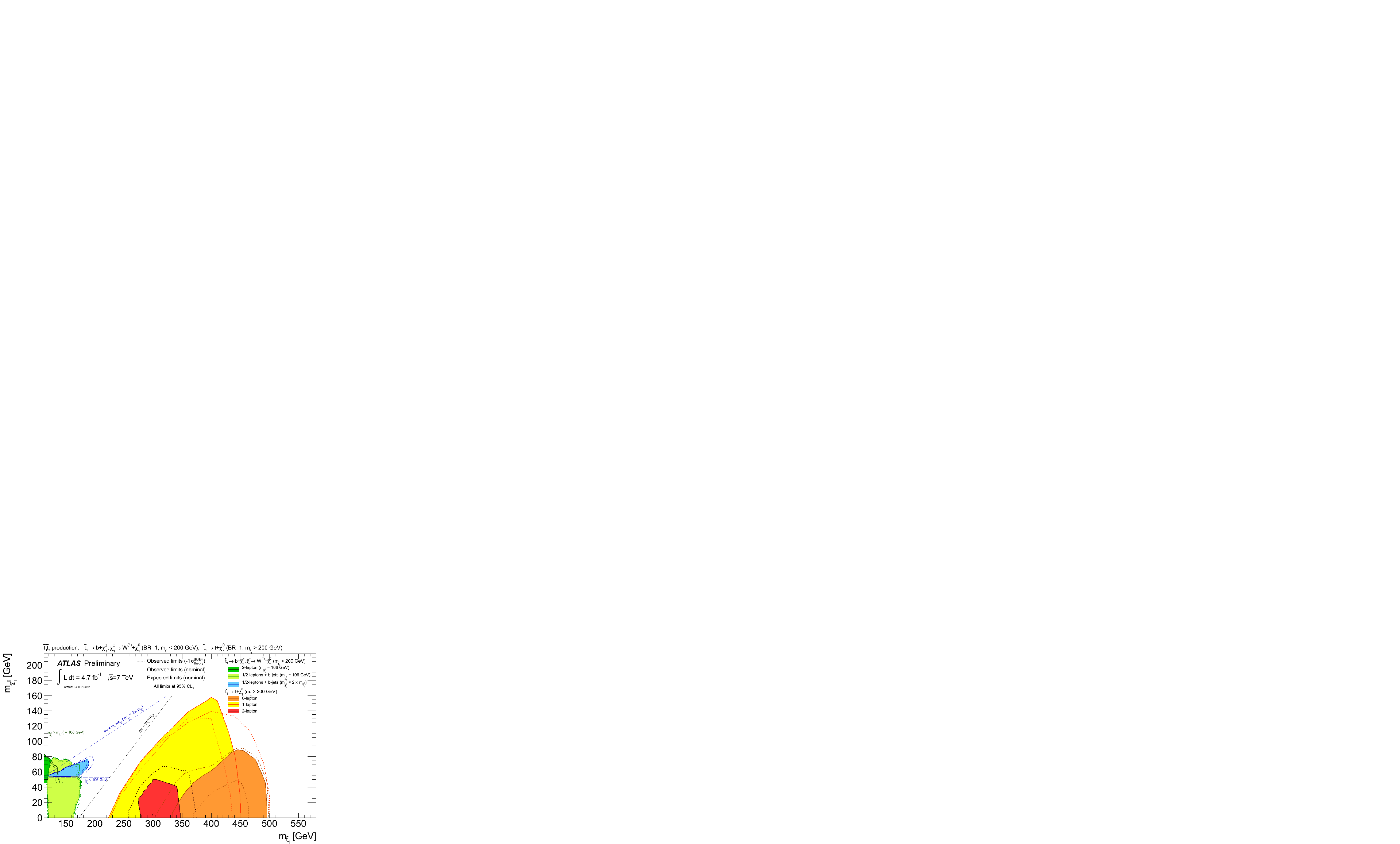}
\caption{An exhibition of the excluded region in the neutralino $\tilde \chi_1^0$ and 
stop mass plane. The low stop mass (white)  region around 200 GeV is not yet excluded. 
Figure  taken from https://twiki.cern.ch/twiki/bin/view/AtlasPublic/SupersymmetryPublicResults.}
\label{fig5}
\end{center}
\end{figure}

The deviations of the MSSM light Higgs boson from its standard model couplings are often 
characterized in terms of decoupling and non-decoupling limits~\cite{Haber:1994mt,Gupta:2012mi}. 
Often  the decoupling limit is defined by  
 $cos^2(\beta -\alpha) <0.05$,
 where $\alpha$ is the 
mixing angle between the two CP-even  Higgses of  MSSM, i.e.,   $H_1^0, H_2^0$ to produce
the mass eigenstates $h^0, H^0$, and is typicallly realized when $M_A  > 300$ ~GeV.
Using the  profile likelihood analysis (see, e.g., Refs. \cite{Feroz:2008xx,Trotta:2008bp})
the  work of Ref.~\cite{Akula:2012kk} gives $m_A> 300$ GeV under the constraint of $m_{h^0} \sim 125$ GeV
so one is in a decoupling phase.
The decoupling phase also explains the experimental limits
from LHCb {\t (http://inspirehep.net/record/1094383)}
on $B_s^0\to \mu^+\mu^-$ being very close to the SM value, since the SUSY contribution proceeds
via the diagram involving the exchange of neutral Higgses  
\cite{Choudhury:1998ze,Babu:1999hn,Bobeth:2001sq,Dedes:2001fv,Arnowitt:2002cq,Mizukoshi:2002gs,Ibrahim:2002fx}.
Thus the most recent result from LHCb gives 
$Br(B^0_s\to \mu^+ \mu^-) < 4.5 \times 10^{-9}$  while the standard model result is 
$Br(B^0_s\to \mu^+ \mu^-) \sim 3.2\times10^{-9}$  implying the supersymmetric contribution to 
this process is small which is consistent with the decoupling limit
(for implications of the $Br(B^0_s\to \mu^+ \mu^-)$ constraint on dark matter see Refs. \cite{Arnowitt:2002cq,Feldman:2010ke}).
In the  analysis carried out in Ref.~\cite{Akula:2012kk}  one finds that while the first two generation squarks are typically heavy several other sparticles  can be relatively light. These include the  gluino, the chargino, the neutralino,
the
stop, the stau. Further, the $\mu$ parameter can be light.  In the top right panel of Fig. (\ref{fig4}) we give
an analysis of the gluino mass $m_{\tilde g}$ vs $\mu$ under the constraint of the 
$\sim 125$ GeV Higgs mass. One finds that there are significant regions of the parameter space (see the lower
left corner of the top right panel of Fig. (\ref{fig4})) where $\mu$ is small, typically $O(200)$ GeV and the gluino
is also relatively light, i.e., typically around a TeV. This region appears to be the prime region for  the
exploration at the next round of LHC experiment. An analysis of signatures for the MSSM Higgs boson in the 
non-decoupling region at the LHC is given in Ref.~\cite{Christensen:2012si}.\\

 Returning to the decoupling limit the deviations of the Higgs  boson couplings to fermions and to vector bosons
  from the SM values are expected to be typically small in the decoupling limit. 
   Thus the couplings of Higgs to the fermions and to the vector bosons  at the tree  level  have the form 

\beqn
L_{h^0A\bar A} = -\frac{m_f}{v} h^0 f\bar f  - \frac{2M_W^2}{v}  h^0 W^{\mu +} W_{\mu}^- - \frac{M_Z^2}{v} 
 h^0 Z^{\mu}Z_{\mu} + \cdots
\eeqn

To test beyond the standard model (BSM) physics one may define the tree level couplings the ratio  
\beqn
R_{h^0A\bar A} = g_{h^0 A\bar A}^{\rm BSM}/g_{h^0 A \bar A}^{\rm SM} \, ,
 \eeqn
 where 
 \beqn
 R_{h^0A\bar A} = 1 + \Delta_{A} \, . 
 \eeqn
 and $\Delta_{A}$ encode the BSM contribution. Thus accurate measurements of $\Delta_{A}$
 are needed to discern BSM physics.  
 How to extract the Higgs bosom couplings from LHC data has been discussed by a number of authors
 (see, for example, Refs. ~\cite{Klute:2012pu,Lafaye:2009vr,Duhrssen:2004cv,Zeppenfeld:2000td}). 
 A post Higgs discovery analysis on the Higgs couplings using LHC data is given in Ref.~\cite{Plehn:2012iz} 
 and exhibited in the bottom left panel of Fig.(\ref{fig4}) where the central values and the errors on the couplings are 
 displayed. The analysis of Ref.~\cite{Plehn:2012iz} finds no significant deviation from the standard model
 in any of the Higgs couplings. However, the errors in the determinations of the couplings are rather
 significant.
One can ask how accurately we  will be able  measure these couplings eventually at the LHC? According to the analysis of Peskin~\cite{Peskin:2012we}
at LHC with $\sqrt{s}=14$ TeV, and ${\cal L} =300$ fb$^{-1}$ one could 
achieve an accuracy in the determination of the Higgs couplings in the range 10-20\% which would 
broadly test the standard model.  
However, it is unlikely that LHC can  achieve an accuracy at the level of  5\% which may be needed to discriminate  Higgs bosons of  new physics models from the Higgs boson of the standard model. 
Better accuracies could be achieved  at the ILC  according to the analysis of Ref.~\cite{Peskin:2012we}.
This is exhibited in the bottom right panel of Fig. (\ref{fig4}).
Thus it is estimated that an ILC with 
1000 fb$^{-1}$ of integrated  luminosity at $\sqrt s=1$ TeV should be able to reach a sensitivity of about 3\% for all the Higgs
couplings~\cite{Peskin:2012we}. Such a measurement will obviously be of considerable value in
the discovery of new physics beyond the standard model.\\

In the standard model the electro-weak corrections to $a_{\mu}=(g_{\mu}-2)/2$ arise from the exchange of 
the $W$ and of  the $Z$ boson. In supersymmetric theories the contribution to $g_{\mu}-2$ can 
arise from the exchange of charginos-sneutrinos and  of  neutralinos-smuons. These contributions at the one loop level
were computed in ~\cite{Yuan:1984ww,Kosower:1983yw}.   Obviously the supersymmetric contribution
depends sensitively on the sparticle spectrum. The  experimental determination of $a_{\mu}$ depends
heavily on the hadronic corrections.  These corrections are estimated by using  $e^+e^-$ 
annihilation cross section and by using the $\tau$ decays. The analysis using $e^+e^-$ annihilation gives
$\delta a_{\mu}= (28.7\pm 8.0)\times 10^{-10} ~(3.6\sigma)$
while for $\tau$-based hadronic contributions one has
$\delta a_{\mu}= (19.5\pm 8.3)\times 10^{-10} ~(2.4\sigma)$~\cite{Hoecker:2010qn,Hagiwara:2011af}.
A more recent analysis using a complete tenth-order QED contribution~\cite{Aoyama:2012wk}
and the hadronic error analysis 
based on $e^+e^-$ annihilation  gives   $\delta a_{\mu}= 24.9 (8.7) \times 10^{-10} ~(2.9\sigma)$ deviation. 
Using one loop supersymmetric contribution to $a_{\mu}$, either experimental result would give some 
tension in SUSY analyses where the sfermion spectrum is heavy. However, 
estimates of two loop effects could reduce this tension~\cite{Heinemeyer:2004yq}.
Alternately part of the Higgs  mass correction could arise from extra matter 
 ~\cite{Moroi:1992zk,Babu:2008ge,Martin:2010dc,Martin:2009bg,Moroi:2011aa,Endo:2011mc }
or through D terms
arising from extra gauge groups under which the Higgs boson could be charged (see, e.g., 
Ref. \cite{Endo:2011gy}). \\

\begin{figure}[t!]
\begin{center}
\includegraphics[scale=0.35]{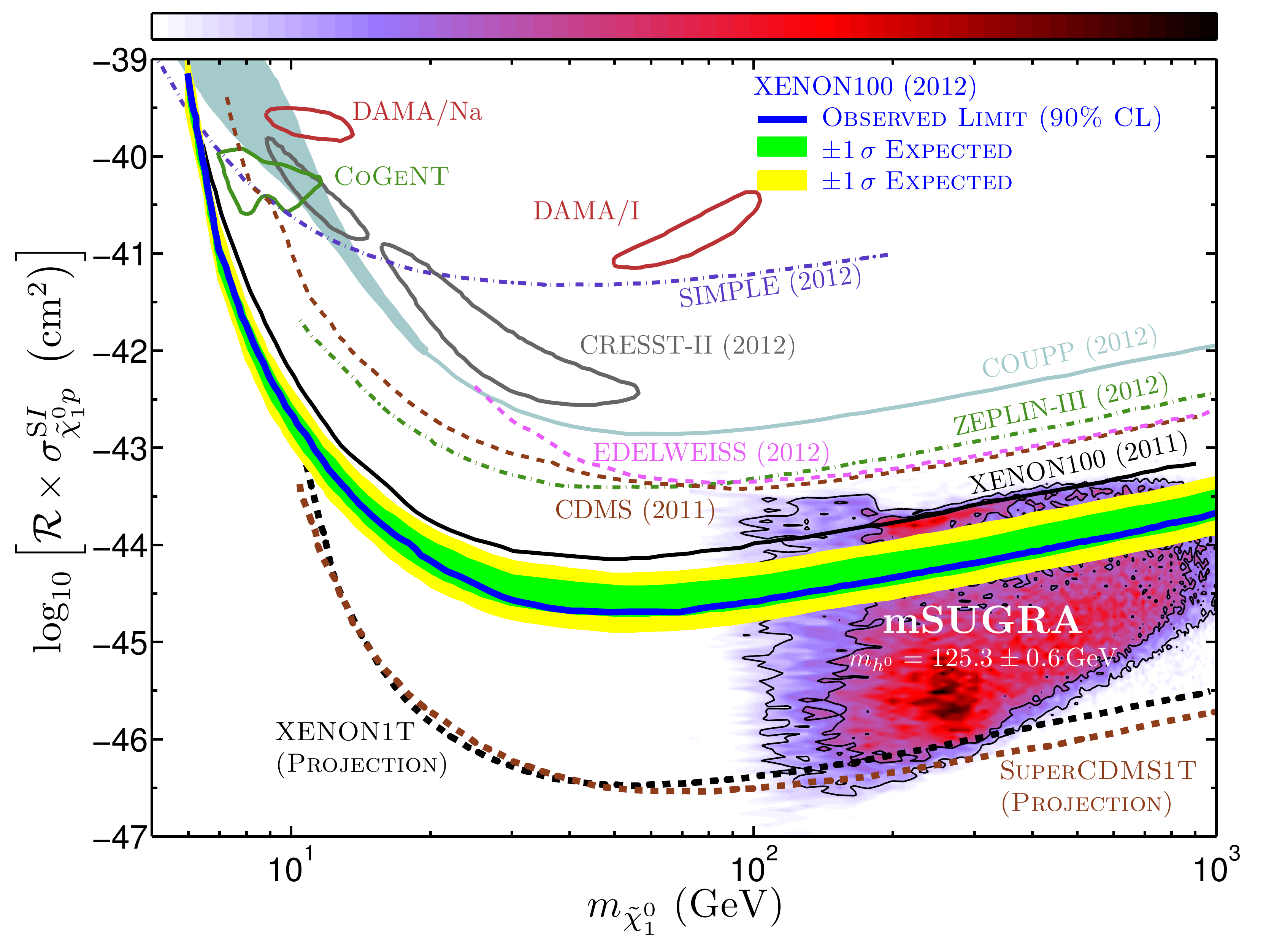}
\caption{
An analysis of the spin-independent neutralino-proton cross section 
exhibited as  a function of the neutralino mass for the model mSUGRA including the
  $\sim 125$ GeV  Higgs mass constraint~\cite{Akulax}. 
 It is seen that most of the parameter space lies
between the current XENON100 limit~\cite{Aprile:2012nq,Aprile:2011hi,Aprile:2011hx,Aprile:2010um} and the limit 
from XENON-1T~\cite{Aprile:2012zx} and from SuperCDMS-1T~\cite{Cabrera:2005zz}.}
\label{fig6}
\end{center}
\end{figure}

\section{Supersymmetry and the cosmic frontier}
Supersymmetry enters in many aspects of the cosmic frontier. These include
(i) dark matter, (ii) dark energy, (iii)  inflation, and (iv)  baryogenesis.
We focus here on dark matter. In supersymmetric theories with R parity, the lightest R parity 
odd particle  (LSP) is stable, and  if such a particle is neutral it can be a  possible candidate for 
dark matter. In SUGRA unified models, over much of the parameter space of models, the neutralino
turns out to be the LSP~\cite{Arnowitt:1992aq}
 and thus the neutralino becomes a candidate for dark matter  with R parity. It is then reasonable to ask
what 
the implications of the discovery of the Higgs boson with a mass of  $\sim 125$ GeV are on dark matter.
It turns out that the $\sim 125$ GeV Higgs mass puts rather severe constraints on the allowed 
parameter space  and thus on the allowed range of the
spin independent cross section~\cite{Akula:2012kk,Akula:2011aa} as exhibited in 
Fig.(\ref{fig6}) for the mSUGRA model. Here one finds that much of the allowed parameter space of the 
model lies within the current XENON-100 limit ~\cite{Aprile:2012nq,Aprile:2011hi,Aprile:2011hx,Aprile:2010um}
and the limit that one will be able to achieve with
XENON-1T ~\cite{Aprile:2012zx} and with superCDMS~\cite{Cabrera:2005zz} in the future.  An analysis of the gaugino-higgsino content of the models analyzed
shows that the  allowed  model points consistent with all constraints have a
significant Higgsino content.  We discuss now another aspect 
of dark matter which relates to cosmic coincidence as discussed below. \\

There is an interesting cosmic coincidence  in that~\cite{Komatsu:2010fb}
\beqn
\frac{\Omega_{DM}h_0^2 }{ \Omega_B h_0^2} = 4.99 \pm 0.20\, ,
\eeqn
where $\Omega_{DM} h_0^2$ is the  dark matter relic density  and $\Omega_B h_0^2$ is the relic density of baryonic 
matter. 
The above appears to indicate that the two are somehow related.  One proposal is that dark matter is 
generated by the transfer of a net  $B-L$ created in the early universe to the dark matter sector.
(For a recent work see ~\cite{Kaplan:2009ag} and for a review see ~\cite{Davoudiasl:2012uw}).
There are two main issues to address for the mechanism to work.  The first concerns 
how to transfer $(B-L)$ generated in the visible sector in the early universe to the 
dark matter sector and the second is how to dissipate thermal dark matter.
Regarding the first one can do so by considering a $B-L$ transfer equation of the form
 $L_{int} = \frac{1}{M^3} \psi^3LH$. 
When the transfer interaction is in equilibrium, one can
solve for  $\Omega_{DM}/\Omega_B$ using the thermal equilibrium
 technique~\cite{Harvey:1990qw,kolbturner}  which gives 
$\frac{\Omega_{DM}}{\Omega_B} = \frac{X\cdot m_X}{B\cdot m_B}$. 
where $X$ is the number of dark matter particles, and $B$ is the number of baryons and
$m_X, m_B$ are their corresponding masses. 
Regarding the second item, i.e.,  thermal dark matter, 
one needs an efficient mechanism for its dissipation. 
Additionally, 
an important issue regarding fermionic dark matter relates to the possibility that such matter 
 can acquire Majorana masses allowing it to oscillate and dissipate.  This issue 
 needs to be addressed as well.

Cosmic coincidence can work in SUSY models with R parity with an extended gauge group.
For example, one may consider  an extension of the conventional SUGRA models with the gauge
group $SU(3)_C\times SU(2)_L\times U(1)_Y$ with an extra $U(1)_X$ gauge factor. 
A convenient choice is $X=B-L$ since $B-L$ is not washed out by spharelon interactions.  
Specifically,  let us suppose we consider a  transfer equation so that 
$ W= \frac{1}{M^3} X^2 (LH_u)^2$.
We assume that the $U(1)_X$ gauge  boson gets its mass via the Stueckelberg mechanism
\cite{Kors:2004dx,Kors:2004ri,Kors:2005uz,Feldman:2006wb,Feldman:2007wj}.
Here there are two dark matter particles, i.e., $\psi_X$ (which carries a lepton number) and 
the conventional dark matter particle of supersymmetry, i.e., the neutralino
$\tilde \chi^0$. Thus the total relic density in this case  is ~\cite{Feng:2012jn}
\beqn
\Omega_{\rm DM} = \Omega_{\rm DM}^{\rm asy} + \Omega_{\rm DM}^{\rm sym}  + \Omega_{\tilde\chi^0}\,.
\eeqn
Cosmic coincidence can work if one assumes that $\psi_X$ is the dominant component (around 90\%) and
the neutralino is the subdominant component (around 10\%). $\Omega_{\rm DM}^{sym}$ can be dissipated by
rapid annihilation  via a Breit-Wigner $Z'_X$ pole.
An interesting possibility to investigate is if the neutralino which is a subdominant component in the relic 
density is still  detectable. 
This appears to be the case in an explicit model constructed~\cite{Feng:2012jn} where  the dark matter carries a gauged quantum number which does not allow Majorana mass for dark fermion, and thus fermionic dark matter does not 
oscillate and is not washed out. 
We note in passing that 
an interesting new development in the analysis of dark matter is the idea of dynamical dark 
matter (DDM) ~\cite{Dienes:2011ja,Dienes:2012yz}.
In the DDM scheme the dark matter in the universe consists   of a large number of components 
which interact with each other and have different masses and abundances~\cite{Dienes:2011ja,Dienes:2012yz}. 
Thus in this case one has multi-component dark matter consisting of an ensemble than than one species. 

\section{SUSY grand unification}
 SUSY $SO(10)$ models, as usually constructed, have two drawbacks, both related to the symmetry breaking sector.  
The first is that  more than one  mass scale  are used in breaking of the GUT symmetry: one to
reduce the rank,  and the other to reduce the symmetry all the way down
to the SM gauge group.
  Thus one uses  $16+\overline {16}$ or $126+\overline{126}$  for rank reduction and uses 
   $45$, $54$ or  $210$ for breaking the symmetry down to the standard model gauge symmetry.
   Finally one uses $10$ plets for breaking the electroweak symmetry.
  Now multiple  step breaking requires  additional assumptions  relating VEVs of the   
different breakings  to explain gauge coupling unification of the electroweak and the 
strong interactions.
A single step breaking does not require such an assumption~\cite{Babu:2006rp,Babu:2005gx}.
The second issue which is generic to grand unified models is the doublet-triplet problem, which
is how to make the Higgs triplets heavy while keeping the Higgs doublets light.  Regarding the
first a single step breaking can be achieved by using $144+\overline{144}$ Higgs to break the GUT symmetry.
Regarding the doublet-triplet splitting there are a variety of ways in which one can  accomplish
this. One of these is the  missing partner mechanism~\cite{Grinstein:1982um,Masiero:1982fe}. 
In SU(5) this is accomplished by 
using a 75 plet of Higgs fields to break the GUT symmetry, while the matter sector consists of 
$5+\bar 5$ of light Higgs fields and a $50+\overline{50}$ of superheavy Higgs fields 
 where $50+\overline{50}$ have the property that they contain one triplet and anti-triplet pair but no doublet pair.  
Allowing for mixings between the light and the heavy heavy Higgs fields one finds that all the triplets and anti-triplets
become heavy while the Higgs doublets remain light ~\cite{Grinstein:1982um,Masiero:1982fe}. 
Recently, one example of  the missing partner 
mechanism in $SO(10)$ was found in Ref.  \cite{Babu:2006nf} which was anchored in
$126+\overline{126}$ of Higgs fields. In Ref. \cite{Babu:2011tw}
 an exhaustive treatment of the missing partner 
mechanism is given. 
Here two additional cases were found which are anchored in $126+\overline{126}$
of Higgs while one unique case is found which is anchored in $560+\overline{560}$~\cite{Babu:2011tw}.
In the latter case one can simultaneously break the gauge symmetry at one  scale and also achieve a 
natural doublet-triplet splitting via the missing partner mechanism.
With models involving  $144+\overline{144}$ and  $560+\overline{560}$ of Higgs fields the analysis of textures is rather
different from the conventional approach (see, e.g., Ref. ~\cite{Chattopadhyay:2001mj}). Here the couplings of the $144$ or 
of $560$ with matter can over only occur at the quartic level which are suppressed by the Planck scale.  While such couplings
can produce the right size of Yukawas for the first and the second generation fermions, one requires additional matter fields
to generate the right size of  third generation masses. In the analysis using $144+\overline{144}$ of Higgs it is 
found ~\cite{Babu:2006rp,Nath:2009nf}
that $b-t-\tau$ unification does not necessarily require a large $\tan\beta$~\cite{Ananthanarayan:1991xp}.  \\

We now briefly comment on the current status of proton stability. Proton decay is an important
discriminant for models based on GUTs and  strings~\cite{Arnowitt:1993pd}.
In SUSY GUTs and in strings
proton decay can proceed under the constraints of R parity via baryon and lepton number
violating  
dimension five and dimension six operators. For dimension six operators the dominant decay
mode of the proton is $p\to e^+\pi^0$ which proceeds via leptoquark exchange
(for reviews see Refs. \cite{Nath:2006ut,Raby:2008pd,Hewett:2012ns}) so that    
 $\Gamma (p \rightarrow e^+ \pi^0) \simeq  \alpha_G^2 \frac{m_p^5}{M_V^4}$ which gives  
$\tau (p \rightarrow e^+ \pi^0) \simeq 10^{36\pm1} yrs$. 
This mode may allow one to discriminate between GUTs and strings. Thus generic D brane models allow only $10^2\bar 10^2$ SU(5) type dimension six operators which give $p\to \pi^0 e^+_L$ while SU(5) GUTs also have $10\bar{10}5\bar 5$ which allow $p\to \pi^0e^+_R$  ~\cite{Klebanov:2003my}.
Additionally the $SU(5)$ GUT models  also give the decay $p\to \pi^+\nu$ which is not allowed in generic D brane models.  However, there exist special regions of intersecting D brane models where the operator $10\bar{10}5\bar 5$ is indeed allowed and the 
 purely stringy proton decay rate can be of the same order as the one from SU(5) GUTs  including the mode $p\to \pi^+\nu$ ~\cite{Cvetic:2006iz}. 
 The current status  of proton decay via baryon and lepton number violating 
dimension six operators is as follows:  Superkamiokande gives the limit~\cite{Abe:2011ts}
 $\tau(p\to e^+\pi^0) > 1.4\times 10^{34} ~{\rm yrs}$ while in the future 
 Hyper-K is expected to achieve a  sensitivity~\cite{Abe:2011ts} of
   $\tau(p\to e^+\pi^0) > 1\times 10^{35}~{\rm yrs}$.
Regarding the proton decay arising from baryon and lepton number
 violating dimension five operators via the Higgsino triplet
exchange the dominant mode is $\bar \nu K^+$ which depends  sensitively  on the sparticle spectrum. 
Proton decay also has a significant dependence on  CP phases~\cite{Ibrahim:2007fb}. 
The current experimental limit from Super-K is $\tau(p\to \bar \nu K^+) > 4\times 10^{33} {\rm yrs}$, while
in the future it is expected that Hyper-K may reach a sensitivity of 
$\tau(p\to \bar \nu K^+) > 2\times 10^{34} ~{\rm yrs}$.
Proton decay from dimension five operators in GUTs and strings in typically too rapid if the sparticle
spectrum is light and one needs a cancellation mechanism or other means to stabilize the proton~\cite{Nath:2006ut}.
Proton stability from these operators is more easily achieved when $m_0$ is large and REWSB
occurs on the Hyperbolic Branch~\cite{Chan:1997bi}.\\

\subsection{Radiative Breaking and R Parity} 
As discussed above
one needs R parity for SUSY dark matter and in GUT models, for proton stability. But R Parity even if preserved at the GUT scale can be broken by RG effects~\cite{Barger:2008wn} (for early works on R parity see Refs.~\cite{Aulakh:1982yn,Khalil:2007dr}). 
Thus suppose R parity arises from a $B-L$ gauge symmetry, i.e.,
$R= (-1)^{2s+ 3(B-L)}$. After spontaneous breaking R parity will be preserved if $3(B-L)$ is an even integer.
However, even if $R$ parity is preserved at the GUT scale, it can be broken by RG effects. Thus consider an extension of MSSM with a $U(1)_{B-L}$  symmetry. Here for anomaly cancellation one needs three right handed neutrino 
fields $\nu^c$. The extended superpotential in this case is
\beqn
W = W_{MSSM} + h_{\nu} LH\nu^c + h_{\nu^c} \nu^c \nu^c \Phi  + \mu_{\Phi} \Phi \bar \Phi \, .
\eeqn
The $B-L$ quantum numbers of the new fields  ($\nu^c, \Phi, \bar \Phi)$ are $(1,-2,2)$. 
A VEV growth for $\nu^c$  will break $R$ parity, but a VEV growth for $\Phi$  won't. 
However, the beta functions due to the coupling of the $\Phi$ to the $\nu^c$  
can turn the mass of $\tilde \nu^c$  tachyonic which leads to a VEV growth for $\tilde \nu^c$ and a violation of $R$  parity.
A Stueckelberg mass growth for the $B-L$ gauge boson (see Sec.(\ref{st}) for a review of the Stueckelberg mechanism)
will preserve $R$ parity since there 
no $\Phi, \bar \Phi$  fields are needed to break the symmetry~\cite{Feldman:2011ms}.

\section{Stueckelberg extensions\label{st}}
In the area of string phenomenology one may classify the current work broadly in the following areas:
(i) String model building which includes looking for string vacua which have the standard model gauge group,
have three generations of quarks and leptons, solve the doublet-triplet problem, and have no exotics.
Desirable models should also have a mechanism for generating soft terms and a mechanism for understanding
the smallness of the cosmological constant;  
 (ii) Investigation of model independent signatures in low scale
compactifications, e.g., via scattering amplitudes; (iii) 
Investigation of other generic or  model independent features: This includes
investigation of hidden sectors which are endemic in string models. Here there is considerable 
work on the investigations of models where communication between the visible sector and the hidden
sector occurs either via gauge kinetic energy mixing or from mixings via the Stueckelberg mechanism. 
 We will focus on the last item, i.e., the implications of the hidden sectors in string theory for 
 particle physics.  Hidden sectors first arose in gravity mediated breaking of supersymmetry where
 supersymmetry is broken in a hidden sector and communicated to the visible sector by gravitational
 interactions. 
 However, the  hidden sector may
 play a larger role. We illustrate this by an example where the communication between the hidden sector
 and the visible sector takes place not by gravity but by a gauge field via the Stueckelberg mechanism.
 Now the \st is quite generic in string compactifications and in the reduction of the higher dimensional theories
 to four dimensions. As an example consider terms in the Lagrangian of the form
  \beqn
  L_1 = -\frac{1}{12} H^{\mu\nu\lambda} H_{\mu\nu\lambda} + \frac{m}{4} \epsilon^{\mu\nu\rho\sigma} 
  F_{\mu\nu} B_{\rho\sigma} \, ,
  \label{L1}
  \eeqn
where $F_{\mu\nu}$ is the field strength of a $U(1)$ gauge field $A_{\mu}$ and $B_{\mu\nu}$ is 
anti-symmetric one form field, and $H^{\mu\nu\lambda}$ is the corresponding field strength
given by $H_{\mu\nu\rho} = \partial_{\mu}B_{\nu\rho}+ \partial_{\nu} B_{\rho \mu}  +  \partial_{\rho} B_{\mu\nu}$.
A simple set of field redefinitions reduce $L_1$ to the form 

\beqn
L_2= -\frac{m^2}{2} (A_{\sigma}+ \partial_{\sigma} \sigma)^2 .
\label{L2}
\eeqn 
where $\sigma$ is related to $H^{\mu\nu\rho}$ so that $H^{\mu\nu\rho} = -m \epsilon^{\mu\nu\rho\sigma}
(A_{\sigma} + \partial_{\sigma} \sigma)$ .

\begin{figure}[t!]
\begin{center}
\includegraphics[height=4cm,width=5cm]{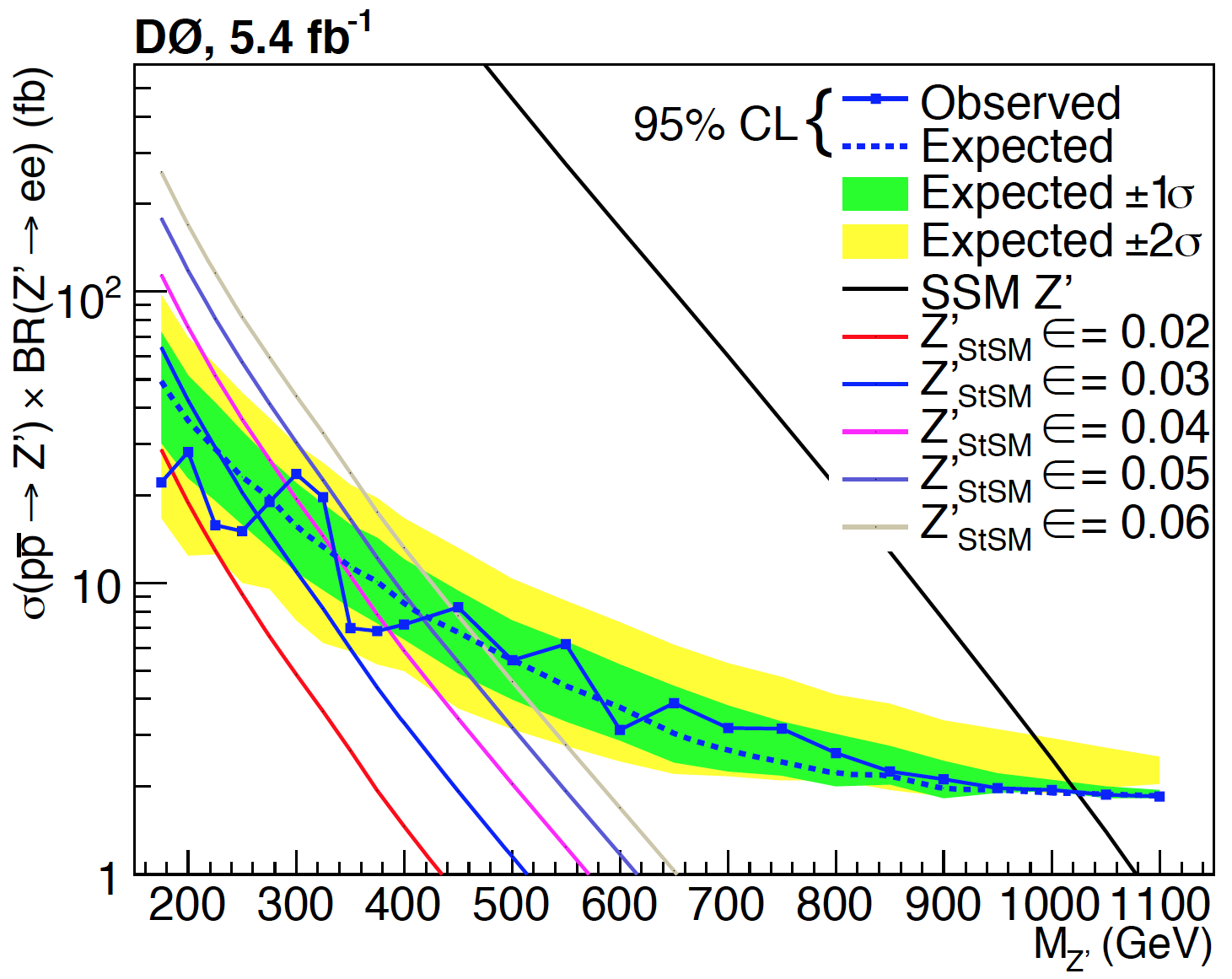}
\includegraphics[height=4cm,width=5cm]{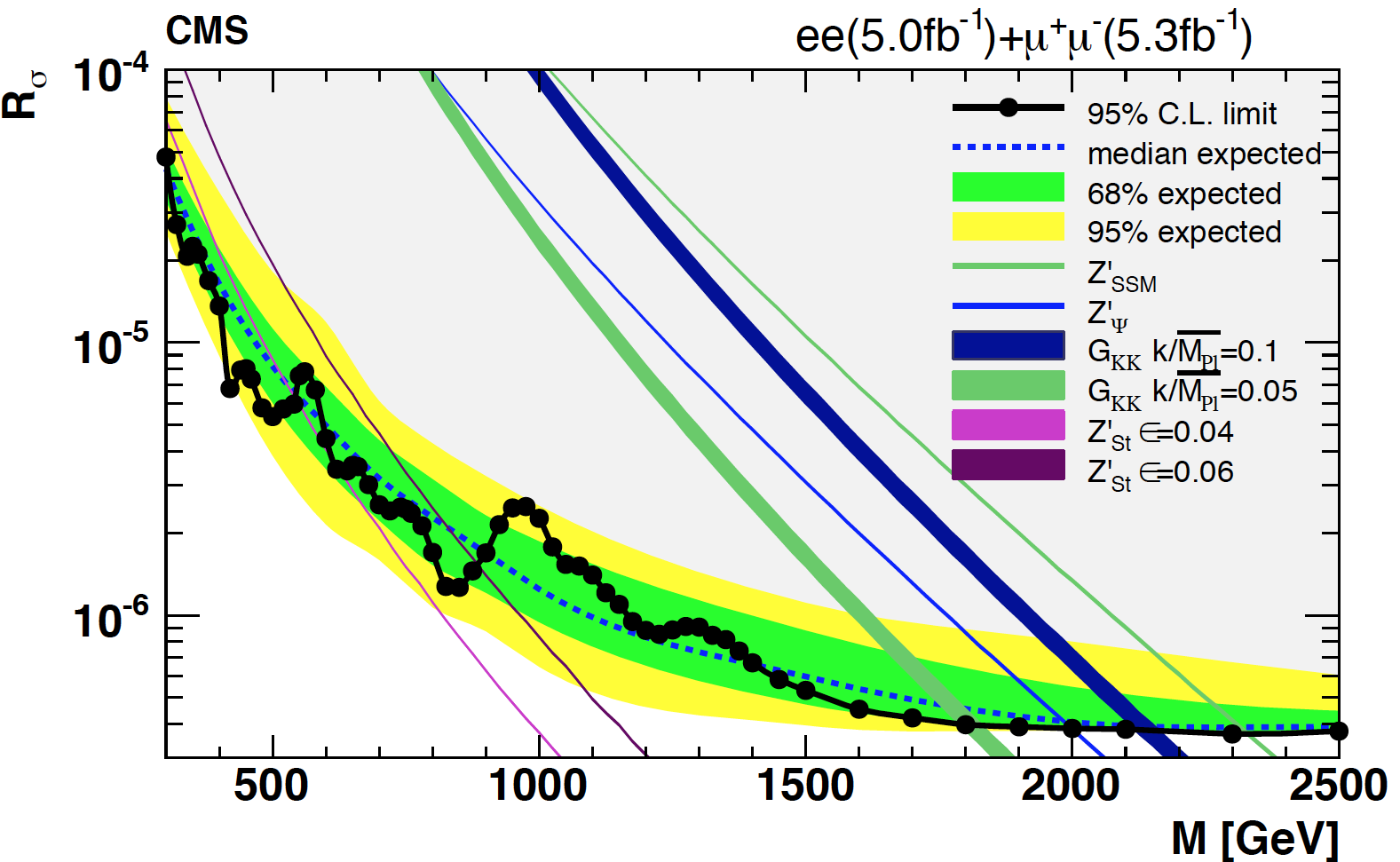}
\caption{Left panel: The constraint on the $Z'_{\rm St}$ mass vs 
 the mixing parameter $\epsilon$
 in the Stueckelberg extension of the
standard model and of the minimal supersymmetric standard model  by the D\O\  Collaboration with 
$5.4$ fb$^{-1}$ of data at the Tevatron   (Taken from Ref. \cite{Abazov:2010ti}). 
 Right panel: The constraint on the $Z'_{\rm St}$ mass vs 
  $\epsilon$  by the CMS Collaboration using 5.0 fb$^{-1}$ of  data for the $e\bar e$ mode and $5.3$ fb$^{-1}$ of data for the $\mu\bar \mu$ mode  (Taken from Ref. \cite{Chatrchyan:2012it}).
One finds that a Stueckelberg  $Z_{\rm St}'$ as low as a few hundred GeV is  still possible.}
\label{fig7}
\end{center}
\end{figure}

Eq.(\ref{L1}) is a very simple form of terms that arise in string theory. 
Thus one finds that the \st arises quite naturally in strings.  Let us assume now that one has 
a hidden sector endowed with a $U(1)_X$ gauge group under which the standard model 
particles are all neutral. Normally there would be no communication between the standard model
sector and the hidden sector in this case. Suppose, however, there was a small mixing between
the hidden and the visible sector via either the kinetic energy sector or via the Stueckelberg sector. 
Thus suppose we have a mixing between the hypercharge field $B_{\mu}$ and the $U(1)_X$ gauge
field $C_{\mu}$ so that the Stueckelberg interaction takes the form ~\cite{Kors:2005uz,Kors:2004iz}
\beqn
L_{St} = -\frac{1}{2} (\partial_{\mu}\sigma + M_1 C_{\mu} + M_2 B_{\mu})^2 .
\label{Lst}  
\eeqn
Combining it with the Higgs mechanism one finds that the vector boson sector now becomes 
a $3\times 3$ matrix rather than a $2\times 2$ matrix as in the case of the 
standard model and for the MSSM~\cite{Kors:2005uz,Kors:2004iz,Kors:2004dx,Cheung:2007ut,Feldman:2007nf,Feldman:2006wb}.
As a consequence of Eq.(\ref{Lst}) the gauge boson of the hidden sector will couple with the
quarks and the lepton of the visible sector and the coupling will be determined by the ratio 
$\epsilon = M_2/M_1$ and such a $Z'$ can be constrained by experimental data 
 (for other related works see, e.g., ~\cite{Cassel:2009pu,Mambrini:2009ad}).  
The current constraints on $\epsilon$ 
arising from the D\O\   Collaboration~\cite{Abazov:2010ti}
  and the CMS detector Collaboration ~\cite{Chatrchyan:2012it} are exhibited in Fig.(\ref{fig7}).
The analysis of D\O\ Collaboration~\cite{Abazov:2010ti} (left panel) was based on $5.4$ fb$^{-1}$ of data and shows that
the $Z'_{\rm St}$ mass could be as low as $\sim 200$ GeV for $\epsilon =0.06$. The  analysis of CMS 
Collaboration~\cite{Chatrchyan:2012it} (right panel) 
was based on $5.0$ fb$^{-1}$ of date for the $e\bar e$ channel and $5.3$ fb$^{-1}$ of data for the
$\mu^+\mu^-$ channel. Here one finds that the $Z'_{\rm St}$ can be as low as $\sim 600$ GeV at 95\%
CL  for the case $\epsilon =0.06$. 

\begin{figure}[t!]
\begin{center}
\includegraphics[height=6cm,width=6cm]{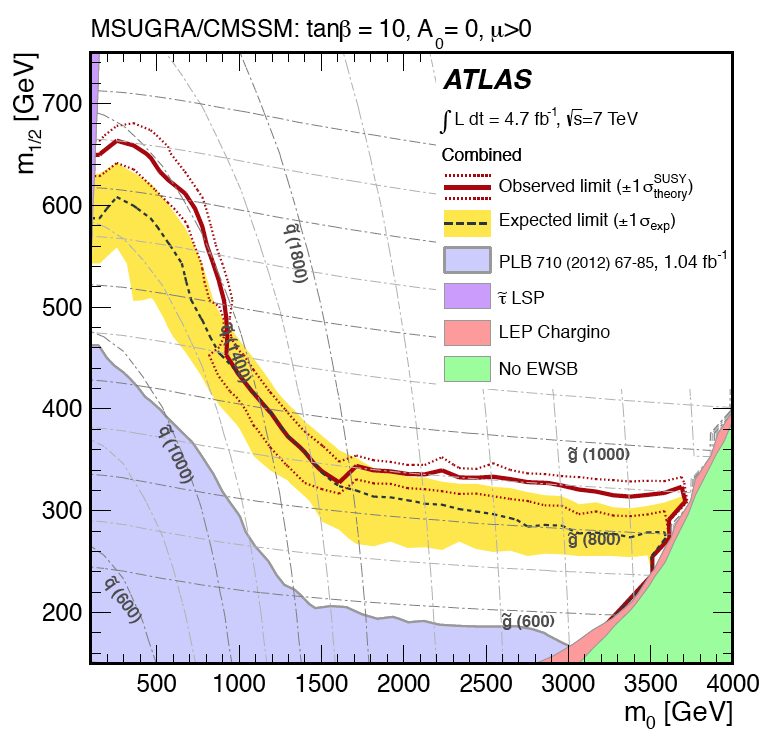}
\includegraphics[height=6cm,width=6cm]{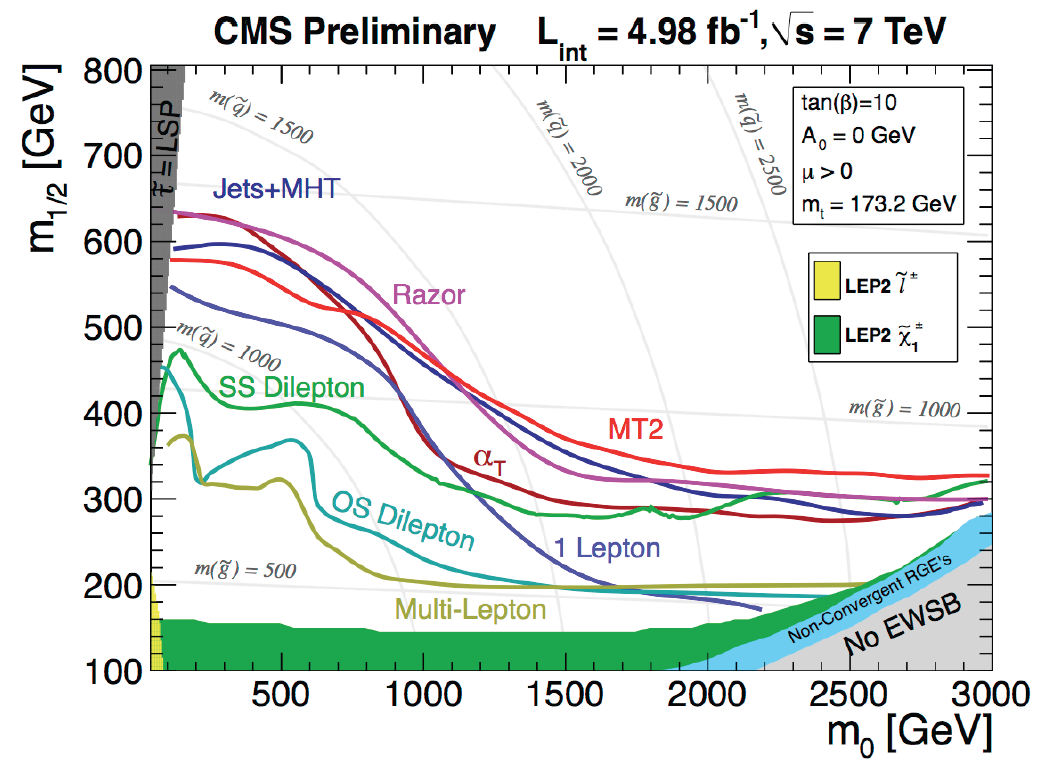}
\caption{The left panel: The reach plot in the $m_{1/2}-m_0$ plane by the ATLAS Collaboration 
with $4.7$ fb$^{-1}$ of integrated  luminosity (taken from Ref. ~\cite{:2012rz}).
Right panel: The reach plot in the $m_{1/2}-m_0$ plane by the CMS Collaboration at 
$\sqrt s=7$ TeV with  $4.98$ fb$^{-1}$ of integrated luminosity (taken from Campagnari's 
talk at SUSY2012~\cite{campagnani}).}
 \label{fig8}
\end{center}
\end{figure}

\section{Prospects for SUSY}
If one believes  that 
the Higgs boson is elementary then unless  one  is willing to live with a large fine tuning, 
supersymmetry is the clear alternative. However, the $\sim 125$ GeV Higgs mass  puts 
a severe constraint on supersymmetric
models. This is because a large loop correction which lifts the tree level Higgs mass of below $M_Z$  to 
$\sim 125$ GeV
 is needed and not every model can support such a large correction.
 Consequently several models  get eliminated while others are severely constrained ~\cite{Arbey:2012dq}.
 The mSUGRA model as well as the non-universal SUGRA models  can lift the tree level Higgs
mass 
 to what is
observed. Indeed in mSUGRA the Higgs boson mass is predicted to lie below $\sim 130$ GeV and it is 
interesting that the experimental value ended up below 
this limit~\cite{Akula:2012kk,Akula:2011aa,Arbey:2012dq}.
The way one generates a large correction to the Higgs boson mass is to have a large $m_0$ which tends to
make the first two generation sfermions heavy. However, it turns  out that in this high scale model there
exist very significant regions of the parameter space where $\mu$ is small, i.e., as low as $\sim 200$ GeV
as well as the 
 gluino, charginos and neutralinos can be light essentially limited by the experimental lower 
limits. 
Thus  the most likely candidates for discovery at the LHC are 
  the    gluino, the  chargino, the neutralino and some of the third generation sfermions.
   The observation of the Higgs boson provides a  strong argument to proceed directly to $\sqrt s=14$ TeV to discover SUSY. 
  In summary  the discovery of the Higgs boson is good news  for supersymmetry. 
A full fledged search for supersymmetry is now desirable.    
The current exclusion plots in the $m_0-m_{1/2}$ plane by  the ATLAS and CMS Collaborations  are given in Fig. (\ref{fig8}). 
       A recent analysis for the reach of LHC at $\sqrt s =14$ TeV 
   is given in Ref. \cite{Baer:2012vr}  and the analysis shows that the discovery reach at the LHC
   at $\sqrt s= 14$ TeV with an integrated luminosity up to 3000 fb$^{-1}$ will be extended by significant 
   amounts.\\
 
 \section{The way forward}
  After the historic results from CERN this year,  there are  clearly two parallel paths that need to be pursued: the first relates to determining 
  more accurately
  the properties of the discovered boson and the second  to a renewed effort to discover
   supersymmetry. Regarding the newly discovered boson    
   the way forward is clear. Although there is little doubt that the newly discovered boson is the long sought after
   Higgs boson,  one must 
 fully  establish its spin and CP properties.  Second, one needs to measure accurately the couplings of the new boson with quarks, leptons and vector bosons.  An identifiable  deviation from the standard model Higgs
 prediction would be a guide to the nature of new
  physics beyond the standard model. However, accurate measurement of the Higgs boson couplings
  are difficult and as mentioned earlier  such measurements at the LHC  
  at best could be at the 10-20\% level by some estimates~\cite{Peskin:2012we}.  
   In view of that it is   timely 
   to start thinking about Higgs factories, i.e.,  
   machines beyond the LHC era,  where such tasks could be better accomplished.
   Aside from ILC or  CLIC, other possibilities  discussed are the   muon collider ~\cite{Barger:2001mi} and more 
   recently a $\gamma \gamma$ Higgs factory SAPPHiRE (Small Accelerator for Photon-Photon Higgs production using Recirculating Electrons)~\cite{Bogacz:2012fs}.
  Aside from the accurate measurement of the properties of the new boson, the main focus now ought to shift to 
the discovery of supersymmetry.

As mentioned  already if one believes in the elementarity of the Higgs boson then supersymmetry is the  most desirable solution.  The analysis including the constraint that
the Higgs boson mass be $\sim 125$ GeV, requires that in SUGRA model with universal boundary conditions, the scalar
mass $m_0$,  is large which makes the first two generations of  sfermions heavy, i.e., of size several TeV.
However, some of the third generation sfermions can be light, specifically, the stop and the stau, due to large
trilinear couplings. Additionally the gauginos, i.e., the gluino, the chargino, and the neutralino can be light
and these could be detected at higher energy and with increased luminosity. It is also possible that the 
light sparticles may have been missed in LHC searches for a variety of reasons. 
Thus, for example, if the mass gap between the LSP and the other low lying sparticles is small, the decay
products of sparticles may not pass the conventional LHC cuts allowing SUSY to remain undetected. 
Alternately without R parity the conventional missing energy signals would not apply. 
Thus, more comprehensive signature analysis are called for exhausting the allowed possibilities in the sparticle 
landscape~\cite{Feldman:2007zn,Nath:2010zj}.
Thus the possibility that  several  sparticles can be relatively light consistent with a $\sim 125$ GeV Higgs mass
is promising for SUSY discovery with further data at the LHC. 

\section*{Acknowledgments}
This  work is  supported in
part by the U.S. National Science Foundation (NSF) grants 
PHY-0757959 and  PHY-070467 and DOE  NERSC grant DE-AC02-05CH11231.

%
\end{document}